\documentclass[11pt,a4paper]{article}

\usepackage{placeins}
\usepackage[margin=0.8in]{geometry}
\usepackage{setspace}
\usepackage{indentfirst}
\usepackage[title,titletoc]{appendix}
\usepackage[english]{babel}
\usepackage{etoolbox}
\usepackage{subcaption}

\apptocmd{\thebibliography}{\setlength{\itemsep}{0pt}}{}{}

\usepackage{amsmath, amssymb, amsfonts, mathtools}
\usepackage{bm}       

\usepackage{graphicx}
\usepackage{grffile}  
\usepackage{subcaption}
\usepackage{float}
\usepackage{wrapfig}
\usepackage[labelfont=bf,labelsep=colon,font=small]{caption}

\usepackage{booktabs}
\usepackage{multirow}

\usepackage[dvipsnames]{xcolor}
\usepackage[
  colorlinks=true,
  linkcolor=MidnightBlue,
  citecolor=MidnightBlue,
  urlcolor=MidnightBlue
]{hyperref}
\usepackage{authblk}
\usepackage[numbers,sort&compress]{natbib}

\usepackage{microtype}

\numberwithin{equation}{section}

\title{\textbf{Dark Matter Induced Scalarization as a Possible Solution to the Hyperon Puzzle}}
\author[1]{Suchana Adhikari}
\author[1]{Teruaki Suyama}
\affil{Department of Physics, Institute of Science Tokyo, Japan
}
\date{}

\begin{document}

\maketitle
\begin{abstract}
We investigate the properties of neutron stars when a massive scalar field, which could comprise all dark matter, is non-minimally coupled to the Ricci scalar. This coupling generates additional contributions to the field’s effective mass, leading to tachyonic instabilities inside neutron stars and giving rise to rich phenomenology. Within this framework, we obtain neutron-star configurations with maximum masses exceeding 2 $M_\odot$, even when hyperons, which typically soften the equation of state and significantly lower the maximum mass, are included. Furthermore, we find that larger coupling strengths lead to multiple solutions for the scalar-field configuration. We analyze the structure of the corresponding effective potential responsible for this behavior. We also investigate how the inclusion of a scalar self-interaction term, in addition to the non-minimal coupling, modifies the resulting neutron-star properties.

\end{abstract}
\section{Introduction}

The bulk properties and internal constitution of neutron stars depend critically on the equation
of state (EoS) of nuclear matter \cite{Prakash1996}. 
When the internal density of a Neutron star (NS) becomes very high, around $(2-3)\rho_0$ \cite{Vidana2022, Nishizaki2002}, the chemical potential of neutrons becomes sufficiently large that they are expected to decay into $\Lambda$ hyperons, forming new hadronic species. 
The inclusion of these hyperons significantly softens the EoS, 
which can reduce the theoretical maximum mass down to, for instance, 
$M_{\mathrm{max}} = 1.4M_\odot$.
On the other hand, some of the heaviest neutron stars measured to date include PSR~J0348+0432 with a mass of $2.01\pm0.04M_\odot$ \cite{Antoniadis2013} and PSR~J1614--2230 with a mass of $1.97\pm0.04M_\odot$ \cite{Demorest_2010}.
These pulsars are in tension with the above mentioned maximum mass.
This discrepancy between theoretical predictions and observations of $2M_\odot$ stars 
is famously known as the ``hyperon puzzle'' in nuclear physics  \cite{Ignazio2016}.
\par\vspace{0.35em}
In this paper, we investigate the possibility that the hyperon puzzle is not a problem of nuclear physics, but rather a consequence of the existence of dark matter.  
It is known that if the mass of dark matter is smaller than $\sim \text{eV}$, 
it must be a boson and behaves as a classical wave whose frequency is characterized by its mass.
Recently, such light dark matter candidates have been extensively discussed 
as a compelling alternative to the traditional WIMP paradigm, 
motivated by the lack of definitive signals from WIMPs \cite{Hui:2021tkt}.
\par\vspace{0.35em}
The purpose of this work is to clarify how the mass and radius of neutron stars in a model 
with a scalar field dark matter $\phi$ coupled to the Ricci curvature 
differ from those in General Relativity (GR).
The basic idea behind this motivation is as follows. 
In the presence of a non-minimal coupling $\frac{1}{2} \xi R \phi^2$, 
the square of the effective mass of the scalar field is given by 
$m_{\rm eff}^2 = m^2 - \xi R$, where $m$ is the mass of the scalar field.  
In the approximation where $R$ is replaced by the energy-momentum tensor
of matter through the Einstein equations ($R = -8\pi G T^\mu_\mu$), 
the effective mass squared inside non-relativistic matter with density $\rho$ is estimated 
as $m^2 - 8\pi G \xi \rho$. 
Therefore, if the density exceeds a critical value $\rho_{\rm crit} = m^2 / (8\pi G \xi)$, 
the trivial solution $\phi = 0$ becomes unstable, and $\phi$ takes a non-zero value which
is the stable point. 
This transition of the scalar field value is known as spontaneous scalarization \cite{Doneva:2022ewd}. 
Since the effective gravitational constant is given by 
$G_{\rm eff} = G / (1 + 8\pi G \xi \phi^2)$, 
the onset of spontaneous scalarization weakens the gravitational strength. 
Consequently, if the critical density for the spontaneous scalarization is lower 
than the saturation density of nuclear matter, 
the weakened gravity inside the neutron star can support a higher maximum mass, potentially exceeding the measured masses of massive pulsars even with the inclusion of hyperons.
\par\vspace{0.35em}
The study of spontaneous scalarization within neutron stars has a long history, 
dating back to the work of Damour and Esposito-Farèse in scalar-tensor theories \cite{DEF1993,DEF1996}. 
In their framework, a massless scalar field coupled to Standard Model particles via a non-trivial conformal coupling triggers the spontaneous scalarization, 
resulting in a significant increase of the maximum mass compared to GR. 
However, embedding such models consistently within cosmology is problematic: 
throughout the evolution, a massless scalar generically acquires a large value inconsistent with present-day observations.
A viable resolution was proposed in \cite{Chen_Suyama_Yokoyama_2015}, where giving the scalar field a mass prevents 
this cosmological blow-up. 
Once the Hubble parameter becomes comparable to the scalar mass, 
the field begins oscillating and $\phi=0$ becomes a cosmological attractor. 
Effects of massive scalar fields on neutron stars with the non-trivial conformal coupling have been investigated in works such as \cite{MorisakiSuyama2017, Ramazano2016}, where it was found that the maximum mass of a neutron star generally
becomes larger than that in GR.
\par\vspace{0.35em}
Although it is possible to transform the actions adopted in the previous studies \cite{Staykov2018, Odintsov2021}
to the Jordan frame where the matter sector is coupled to the metric only, the resulting Lagrangian for the metric and $\phi$ takes a very complicated form in general.
On the other hand, in this work, we assume the scalar field has, in addition to the mass term and the self-interaction term, only the simple non-minimal coupling term to gravity in the Jordan frame.
Reference \cite{Arapogulu2019} is similar to our work in that it investigates the structure of neutron stars with a non-minimal coupling to the Ricci scalar in the Jordan frame. 
However, the study focused on the massless case and adopted a Higgs-like potential, where $\phi$ is unstable even in the vacuum. 
In this paper, we focus on the case where the instability is specifically induced by a massive scalar field in the high-density environment inside the star.
\par\vspace{0.35em}
A similar model has also been considered in \cite{Degollado2024}, where a massive scalar field non-minimally coupled to gravity through the Ricci scalar was studied using a different numerical approach. While qualitatively similar behavior was reported, that work considered polytropic equations of state. The focus of this paper, however, is to test a possible resolution of the hyperon puzzle by employing hyperonic equations of state, in contrast to the polytropic equation of state adopted in the previous work.
\par\vspace{0.35em}
This paper is organized as follows. In Section~2, we present the action and equations of motion governing the system 
and discuss the profile of the scalar field inside the neutron star qualitatively
based on the effective potential.
Section~3 describes the numerical scheme used to obtain solutions. Section~4 contains our results, including the influence of the coupling parameters on mass–radius relations and scalar profiles, and their physical implications. Finally, we summarize our findings and discuss observational consequences in Section~5.

In what follows, we consider the geometrized units corresponding to $c=G=1$.

\section{Basic equations}\label{secbasic}

\subsection{Action of our model and covariant field equations}
In this subsection, we give the action of our model considered in this paper. 
As mentioned in the Introduction,
we assume that dark matter consists of a light scalar field behaving as wave-like dark matter 
with non-minimal coupling to gravity through the Ricci scalar. 
The action of the system is defined as
\begin{equation}
S[g_{\mu \nu}, \phi,A] =\int d^4x\sqrt{-g}\Bigl(\frac{R}{16\pi}-\frac{1}{2} (\nabla\phi)^2-\frac{1}{2} 
m^2\phi^2+\frac{\xi}{2} R\phi^2 \Bigr)+ S_M [A,g_{\mu\nu}].
\label{action}
\end{equation}
Here, $m=\tilde{m}/\hbar$, $\tilde{m}$ being the mass of the dark matter scalar field $\phi$,
$\xi$ is the dimensionless parameter representing the strength of the non-minimal coupling,
and $S_M$ is the action of baryonic matter symbolically denoted as $A$.
In this paper, we only consider the case where $\xi>0$ because the spontaneous scalarization
happens in this case \footnote{In principle, the spontaneous scalarization
can happen even when $\xi <0$ if $\rho-3P<0$. 
This condition may be achieved at the core of the neutron star where the nuclear matter is extremely 
dense and the sound speed exceeds $1/\sqrt{3}$. 
}.
We assume the matter is not directly coupled to $\phi$.

The gravitational field equations are obtained by taking the variation of the action
with respect to $g_{\mu \nu}$ and they are given by
\begin{equation}
G_{\mu\nu}=\frac{8\pi }{(1+8\pi \xi\phi^2)}\Bigl[T_{\mu\nu}^{(\phi)}-\xi(g_{\mu\nu}\square-\nabla_\mu\nabla_\nu)\phi^2+T_{\mu\nu}^{(M)}\Bigr].
\label{einstein-eq}
\end{equation}
Here we have defined the energy-momentum tensor $T_{\mu\nu}^{(\phi)}$ of $\phi$ as
the one used for the case of minimal coupling;
\[T_{\mu\nu}^{(\phi)}=\nabla_\mu\phi\nabla_\nu\phi-g_{\mu\nu}\Bigr(\frac{1}{2}
\nabla ^\sigma\phi\nabla_\sigma\phi+\frac{1}{2} m^2\phi^2\Bigr).
\label{scalar-emtensor}\]
$T_{\mu\nu}^{(M)}$ is the energy-momentum tensor of matter which we treat as perfect fluid
and it is given by
\begin{equation}
T_{\mu \nu}^{(M)}=(\rho+p) u_\mu u_\nu +pg_{\mu \nu}.
\end{equation}
The scalar field equation is obtained by the variation with respect to $\phi$
and it is given by
\begin{equation}
\square\phi-m^2 \phi+\xi R \phi =0. \label{equation-phi} 
\end{equation}

One important feature that can be immediately read from the equation (\ref{einstein-eq}) 
is that at the location of a non-vanishing value of $\phi$, the gravitational constant is effectively
shifted to
\begin{equation}
G_{\rm eff}=\frac{1}{1+8\pi \xi \phi^2} <1.
\end{equation}
Thus, the gravitational strength between matter becomes weaker in the spontaneous scalarization phase than in the normal phase.

\subsection{The basic equations for the neutron star}
In this paper, we investigate a static, spherically symmetric neutron star configuration 
within the framework defined by Eq. (\ref{action}). 
Our primary objective is to determine how the maximum mass of these stars depends on the free parameters $m$ and $\xi$.
While neutron stars in reality rotate, rendering the assumption of spherical symmetry 
an approximation, 
previous studies \cite{Konstantinou2022} indicate that even at a rotational frequency of $350$ Hz, 
the total mass fluctuates by only ${\cal O}(1\%)$. 
In the absence of a dedicated quantitative study on rotational effects within our specific model, 
we proceed under the assumption that such effects remain similarly negligible.
Furthermore, the static condition imposed on the scalar field $\phi$ is an approximation; 
in a cosmological context, $\phi$ oscillates with an angular frequency $m$ 
to account for dark matter. 
However, since the amplitude of these oscillations is expected to be significantly 
suppressed relative to the field value within the NS, 
we anticipate that their impact on the overall stellar structure is minimal. 
A rigorous quantitative evaluation of these oscillations remains beyond the scope of this work. Consequently, we assume a boundary condition where $\phi$ asymptotically decays to zero 
at large distances from the neutron star.
\par\vspace{0.35em}
Under the above assumptions, without a loss of generality, the metric of the system
under consideration can be written as
\begin{equation}
ds^2=-e^{2\alpha(r)}dt^2+e^{2\Lambda(r)}dr^2+r^2 (d\theta^2+\sin^2 \theta d\varphi^2 )
\label{metric},
\end{equation}
where $r$ is the area radius and $\alpha (r)$ and $\Lambda (r)$ are unknown functions
that are determined as solutions of the field equations.
With this notation, $T_{\mu\nu}^{(M)}$ becomes
\begin{equation}
T_{\mu\nu}^{(M)}={\rm diag} 
(\rho e^{2\alpha},pe^{2\Lambda},pr^2,pr^2\sin^2\theta).
\label{matter-emtensor}
\end{equation}
Finally, $\phi$ is also a function of $r$ only; $\phi=\phi (r)$.
Plugging these expressions, non-trivial components of the gravitational field equations (\ref{einstein-eq}) are given by

\begin{align}
\frac{e^{2\Lambda}}{r^2}-\Bigr(\frac{1}{r^2}-\frac{2\Lambda'}{r}\Bigr)=\frac{8\pi}{(1+8\pi \xi\phi^2)}\Bigl[\frac{\phi'^2}{2}+e^{2\Lambda}\frac{m^2\phi^2}{2}+2\xi \Bigl((\frac{2}{r}-\Lambda')\phi'\phi+\phi''\phi+\phi'^2\Bigr)+\rho e^{2\Lambda}\Bigr],\label{tteq} \\
\frac{1}{r^2}(1-e^{2\Lambda})+\frac{2}{r}\alpha'=\frac{8\pi}{(1+8\pi \xi\phi^2)}\Bigl[\frac{\phi'^2}{2}-e^{2\Lambda}\frac{m^2\phi^2}{2}-2\xi(\alpha'+\frac{2}{r})\phi' \phi+p e^{2\Lambda}\Bigr], \label{rreq} \\
\alpha''-\alpha'\Lambda'+\alpha'^2+\frac{\alpha'-\Lambda'}{r} 
= \frac{8\pi }{(1+8\pi \xi\phi^2)}\Bigl[
    -\frac{\phi'^2}{2}-e^{2\Lambda}\frac{m^2\phi^2}{2}
    -2\xi\phi'^2
    -2\xi\phi\Bigl((\alpha'-\Lambda'+\tfrac{1}{r})\phi' + \phi''\Bigr)
    + pe^{2\Lambda}
\Bigr].\label{thetathetaeq}
\end{align}
Here, the prime denotes the derivative with respect to $r$.
The equation for $\phi$ (\ref{equation-phi}) becomes

\begin{equation}
\phi''+\Bigr( \alpha'-\Lambda'+\frac{2}{r}\Bigr)\phi'-e^{2\Lambda}m^2\phi+2\xi\Bigl[\frac{e^{2\Lambda}-1}{r^2}-\alpha'^2+\alpha'\Lambda'-\frac{2}{r}(\alpha'-\Lambda')-\alpha''\Bigr]\phi=0.\label{eomeq}
\end{equation}

These four equations, supplemented with the equation of state for the nuclear matter 
which connects $p$ and $\rho$
enable to determine the five unknown functions $(\alpha, \Lambda, \phi, \rho, p)$
under suitable boundary conditions.
\par\vspace{0.35em}
The above equations do not take forms ready for numerical integration.
After mathematical manipulations, these equations are transformed into the following forms
that allow a direct implementation in the numerical code;
\begin{align}
\alpha'=\frac{1}{2 r (1+\eta +8 \pi  \xi  r \phi' 
   \phi)}(e^{2\Lambda}A_1+A_0), \label{D-alpha} \\
\Lambda'=\frac{1}{2r (1+\eta )\left(1+\eta +6\xi\eta\right) (1+\eta +8 \pi  r\xi  \phi'\phi)}(e^{2\Lambda}B_1+B_0), \\
p'=-\frac{(p+\rho)}{2 r (1+\eta +8 \pi  \xi  r \phi' \phi)}(e^{2\Lambda}A_1+A_0), \label{D-P} \\
\phi''=\frac{1}{r(1+\eta)(1+\eta+6\eta\xi)}(e^{2\Lambda}C_1+C_0). \label{D-phi}
\end{align}
Here $\eta = 8\pi \xi \phi^2$.
The coefficients $A_a, B_a, C_a (a=0,1)$ are functions of $(\alpha, p, \Lambda, \phi, \phi')$ 
and do not contain any quantities, which are derivatives of $(\alpha, p, \Lambda, \phi, \phi')$, 
on the left-hand sides.
Explicit expressions of those coefficients are given in Appendix~\ref{appendix-coefficients}.
\par\vspace{0.35em}
The boundary conditions for the functions $(\alpha, \Lambda, p, \phi)$ imposed
to solve equations (\ref{D-alpha})-(\ref{D-phi}) are given by
\begin{equation}
\label{boundary-condition}
\alpha (0)=0,~~~~~
\Lambda (0)=0,~~~~~
\phi'(0)=0,~~~~~
\phi (r\rightarrow\infty)=0,~~~~~
p(0)=p_c.
\end{equation}
The third condition $\phi' (0)=0$ ensures that $\phi$ is regular at the 
center of the NS.
The pressure at the center $p_c$ is a free parameter that can be varied,
and different values of $p_c$ correspond to different mass of the NS.
\par\vspace{0.35em}
As, we will see later, the scalar field profile is such that it vanishes 
outside the NS, but it still has some presence around the neutron star surface. 
Hence, the gravitational mass of the NS considered would 
take into account for the scalar field mass outside the surface of the star.
Once the equations are solved up to a sufficiently large distance 
$r_\infty (\gg 1/m)$ 
where
$\phi$ decays to practically zero, and the spacetime metric settles down
to the Schwarzshild metric,
the mass of the NS, '$M$' is determined by a relation
\begin{equation}
M =\frac{r_\infty}{2} (1-e^{-2\Lambda (r_\infty )}).
\end{equation}
Note that the masses obtained in this work can be directly 
compared to the mass of PSR~J1614--2230 \cite{Demorest_2010}, which is 
based on the Shapiro time delay measurement,
as long as $r_\infty$
is smaller than the distance between the pulsar and the companion star, which is of the order of $10^{7}$ km.

The radius $R_S$ of the NS is determined as the solution of 
\begin{equation}
p(R_S)=0.
\end{equation}

\subsection{Qualitative behavior of the spontaneous scalarization}
\label{Qualitative-behavior}
Before presenting the numerical solutions to equations (\ref{D-alpha})–(\ref{D-phi}) 
under the boundary conditions (\ref{boundary-condition}), 
it is instructive to examine the qualitative behavior of $\phi(r)$. 
Establishing this physical intuition will help us interpret the subsequent numerical results.

\begin{figure}[t]
    \centering
    \includegraphics[width=0.7\textwidth]{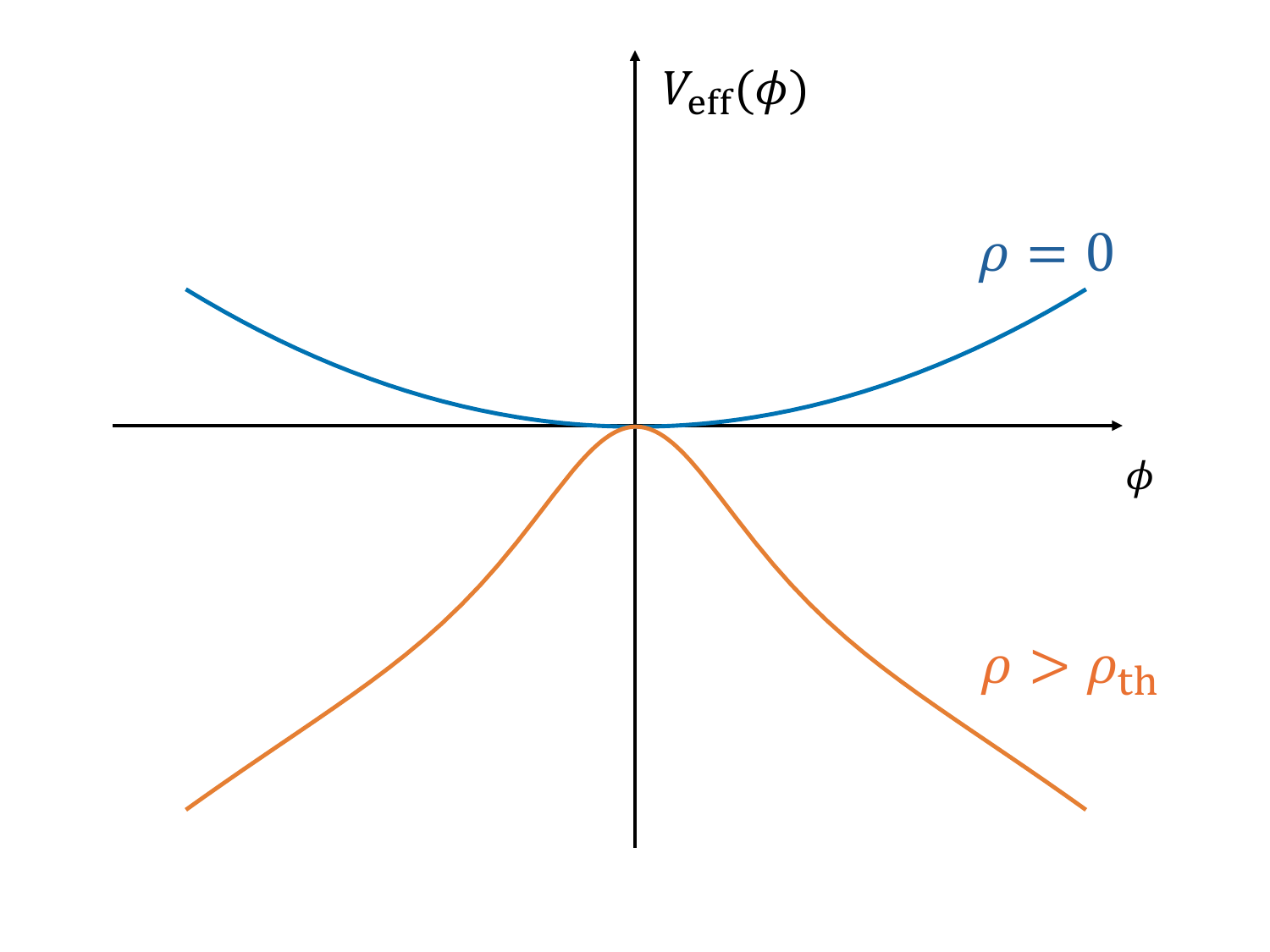}
    \caption{The effective potential in the absence of matter $\rho=0$ (blue curve)
    and in the presence of dense matter $\rho >\rho_{\rm th}$ (orange curve).}
    \label{fig: Effective potential}
\end{figure}

To this end, we use the equation of $\phi$ (\ref{equation-phi}) assuming static and
spherically symmetric configuration;
\begin{equation}
\label{eq-toy-model}
\phi ''+\Bigr( \alpha'-\Lambda'+\frac{2}{r}\Bigr)\phi'-m^2 \phi +\xi R \phi =0.
\end{equation}
An important point is that $R$, which acts as changing the mass term, 
depends on location,
in particular, whether it is inside or outside the NS.
To make this point explicit, we express $R$ in terms of $\phi, \rho$ and $p$ by taking trace of
Eq.~(\ref{einstein-eq}):
\begin{equation}
R=\frac{8\pi}{1+8\pi \xi \phi^2} \left( {(\nabla \phi)}^2+2m^2 \phi^2 +3\xi \Box \phi^2
+\rho -3p \right).
\end{equation}
Ignoring the spatial variation of $\phi$ and making a non-relativistic approximation
($\rho \gg p$) for $R$, Eq.~(\ref{eq-toy-model}) becomes
\begin{equation}
\label{eom-phi-veff}
\phi ''+\Bigr( \alpha'-\Lambda'+\frac{2}{r}\Bigr)\phi'-\frac{dV_{\rm eff}(\phi)}{d\phi} =0,
\end{equation}
where we have introduced the effective potential defined by
\begin{equation}
\label{def-effective-potential}
V_{\rm eff}(\phi)=-\frac{m^2}{2} \phi^2+\left( \frac{m^2}{8\pi \xi}-\frac{\rho}{2} \right)
\ln \left( 1+8\pi \xi \phi^2\right).
\end{equation}
Taylor-expanding this function around $\phi=0$, we have
\begin{equation}
V_{\rm eff} (\phi)=\frac{1}{2} \left( m^2-8\pi \xi \rho \right) \phi^2+{\cal O}(\phi^4).
\end{equation}
In vacuum where $\rho=0$, $V_{\rm eff}\approx \frac{1}{2}m^2\phi^2$ has the shape of a parabola that opens upward around $\phi=0$ (see Fig.~\ref{fig: Effective potential}).
Thus, $\phi=0$ is the stable point.
In the presence of matter, if $\rho$ exceeds the critical value $\rho_{\rm th}=m^2/(8\pi \xi)$,
$V_{\rm eff}$ takes the shape of a distorted parabola that opens downward (see Fig.~\ref{fig: Effective potential}).
Under this condition, $\phi=0$ becomes unstable, which in turn triggers 
spontaneous scalarization.
In terms of the Compton wavelength $\lambda_\phi$, the critical density $\rho_{\rm th}$ becomes

\begin{equation}
\label{critical-density}
\rho_{\rm th}\approx \rho_0  
{\left( \frac{\lambda_\phi}{100~{\rm km}} \right)}^{-2} \xi^{-1},
\end{equation}
where $\rho_0=2\times 10^{17}~{\rm kg/m^3}$ is the nuclear density.
Thus, there is a parameter region where $\rho_{\rm th}$ is smaller than the nuclear density
and the spontaneous scalarization happens inside the NS.

With this crude picture in mind, let us go back to the equation (\ref{eom-phi-veff}).
This equation becomes equivalent 
to the equation of motion of a particle moving in one-dimension
under the influence of the potential $U=-V_{\rm eff}$ and the time-dependent friction 
by identifying $\phi$ with the space coordinate $x$:
\begin{equation}
\label{eom-point-mass}
{\ddot x}+\frac{2}{\tau}{\dot x}=-\frac{dU}{dx},
\end{equation}
where the metric contribution to the friction term has been dropped and 
$\tau (=r)$ has been introduced to emphasize that the radial coordinate
is now interpreted as (fictitious) time variable $\tau$, and ${\dot x}=\frac{dx}{d\tau}$.
The particle starts its motion with zero velocity at the initial time $\tau=0$ and 
reaches $x=0$ in the infinite future $(\tau \to \infty)$.

\begin{figure}[t]
\centering

\begin{subfigure}{0.45\textwidth}
    \centering
    \includegraphics[scale=0.28]{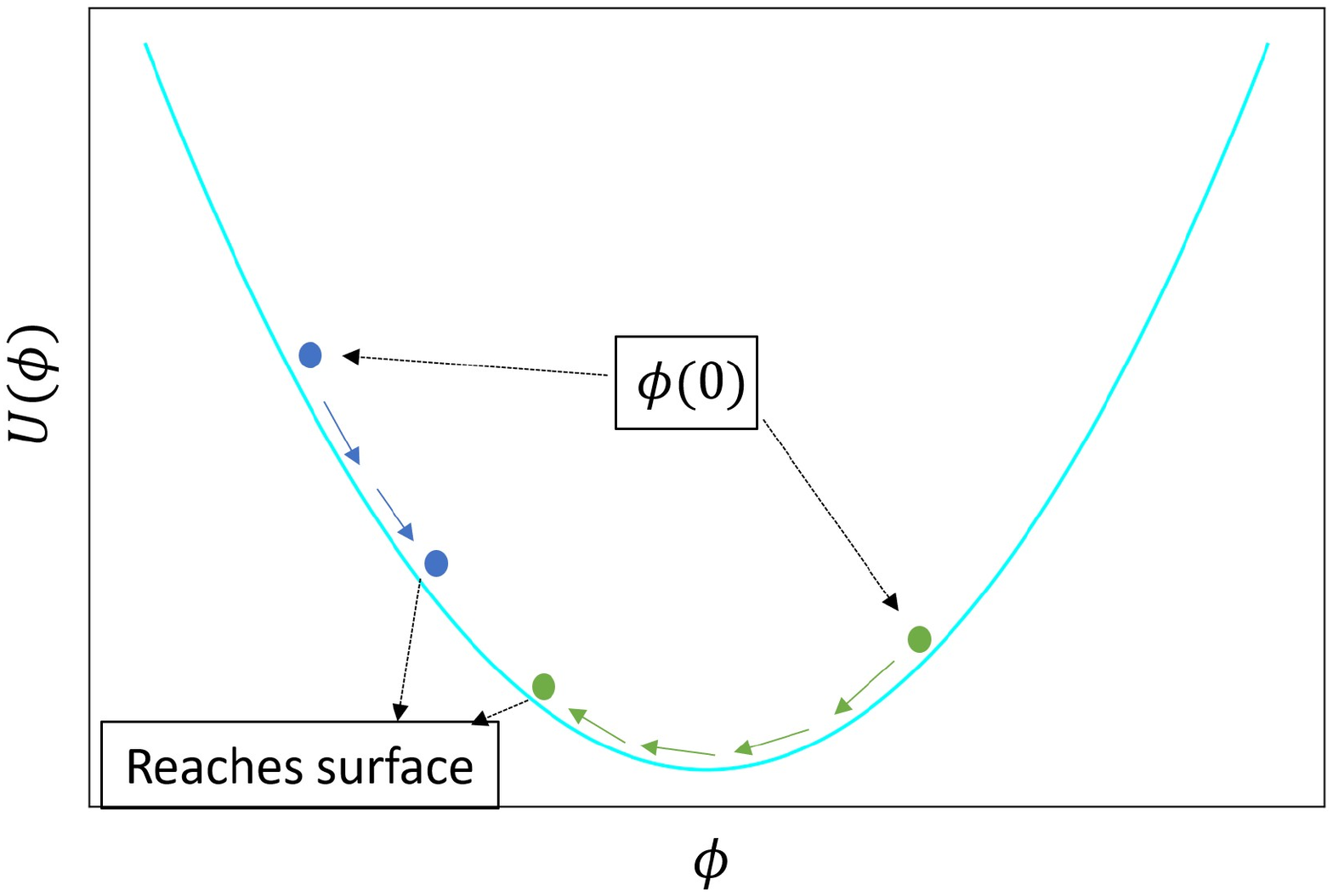}
    \caption{}
\end{subfigure}
\hfill
\begin{subfigure}{0.45\textwidth}
    \centering
    \includegraphics[scale=0.28]{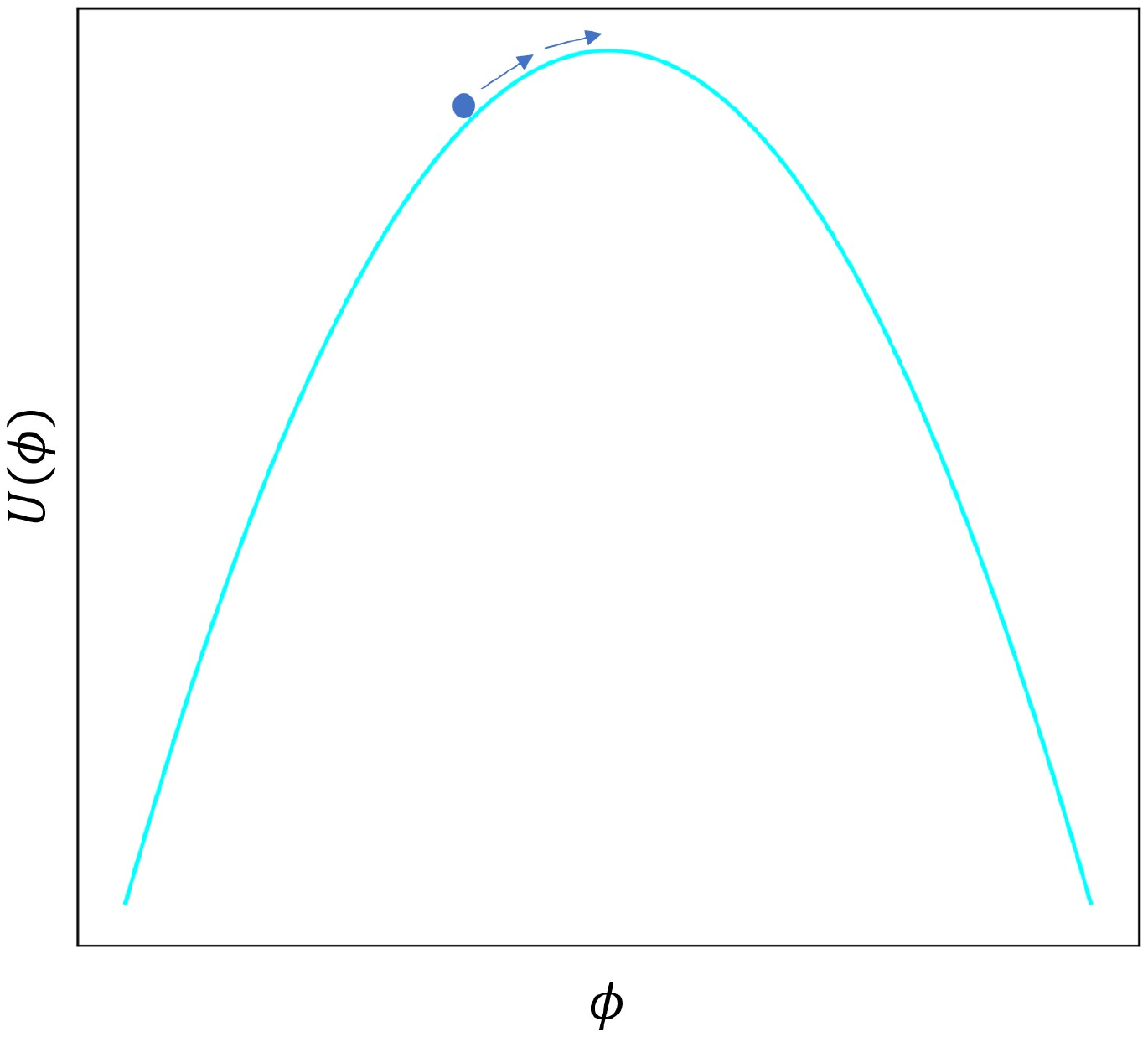}
    \caption{}
\end{subfigure}

\caption{Plot explaining the trajectory of the scalar field particle. (i) The blue ball represents the particle that does not oscillate and is the most stable configuration. The green ball oscillates about the origin before reaching the surface. (ii) The motion of the particle after reaching the surface must reach 0.}

\label{fig: particle trajectory}
\end{figure}

Figure~\ref{fig: particle trajectory} illustrates the behavior of such a scalar field particle whose trajectory corresponds to a non-trivial solution for the equations of motion and obeys the boundary conditions (i.e., decaying outside the star).
Initially, the potential $U$ has a shape given by the left-panel of Fig.~\ref{fig: particle trajectory}
and has a minimum at the origin. 
Thus, the particle initially at rest at position $x(0)$ starts to roll down towards zero.
When $\frac{d^2U}{dx^2}\big|_{x=0}$ inside the NS is small, the potential well around the stable point at $x = 0$ is relatively shallow. 
In this case, the particle typically does not oscillate around the origin. 
Instead, it reaches some value of $x$ at the {\it time} $\tau=R_S$ corresponding to
the stellar surface, 
and thereafter it climbs up the potential (right panel) and gradually approaches zero 
at infinity (blue particle depicted in Fig.~\ref{fig: particle trajectory}).
On the other hand, when $\frac{d^2U}{dx^2}\big|_{x=0}$ inside the NS is large, the potential becomes substantially deeper, providing a stronger effective force and larger acceleration. 
This allows the scalar field to oscillate around the origin before reaching the stellar surface. 
In such cases, multiple non-trivial solutions may exist (green particle depicted in Fig.~\ref{fig: particle trajectory}). 
The most stable configuration corresponds to the ground state, i.e.\ the non-oscillating solution.
In this manner, the spontaneous scalarization can be qualitatively understood in terms of
the classical motion of a particle under the time-dependent potential.

In the small-mass limit, $\frac{dU}{dx}$ in Eq.~(\ref{eom-point-mass}) reduces to
\begin{equation}
\frac{dU}{dx}=\frac{4\pi \xi \rho}{1+8\pi \xi x^2}x^2.
\end{equation}
Consequently, the EoM (\ref{eom-point-mass}) becomes independent of $m$.
Furthermore, all terms containing $m$ in the gravitational field equations vanish smoothly
as $m$ approaches zero.
This suggests that the structure of the NS becomes insensitive to $m$ 
as $m \to 0$.
\FloatBarrier

\section{Numerical methodology}
In the case of standard GR, 
the basic equations describing the structure of a static, spherically symmetric neutron star reduce to a single equation known as the TOV equation \cite{Tolman1939, Oppenheimer1939}.

The numerical solutions for the GR case are widely available in the literature. 
In this section, we present the numerical technique to determine the structure of the NS
and to construct the mass--radius ($M$--$R$) curve
in our model by solving the coupled system of differential equations (\ref{D-alpha})-(\ref{D-phi}), 
which depend on the variables $(p,\rho,\phi,\Lambda)$ and the parameters $(\xi, m)$.
\par\vspace{0.35em}
We begin by selecting an equation of state (EoS) for dense matter. 
In this work, we consider two widely used hyperonic EoS models from the CompOSE database: 
OPGR(GM1Y4)~\cite{Oertel2015} and BHB(DD2$\Lambda$) with electrons~\cite{Banik2014}. 
The OPGR(GM1Y4) EoS describes cold, $\beta$-equilibrated neutron star matter including the full baryon octet. 
The BHB(DD2$\Lambda$) EoS is based on the DD2 relativistic mean-field model and includes $\Lambda$ hyperons along with leptons, yielding a maximum mass of $M_{\rm max}\approx1.95M_\odot$. 
Since we focus on cold neutron stars, the temperature is fixed to $T=0$, or to the lowest temperature available in the CompOSE tables.

Once the EoS is fixed,
for each value of $p$, the corresponding value of $\rho$ is extracted from EoS tables 
obtained from the CompOSE data site. 
The numerical workflow then proceeds as follows:

\begin{enumerate}
    \item For a chosen central pressure $p_c$, we determine the corresponding central density 
    $\rho(r=0)$ from the EoS table. We assume the central field value $\phi(r=0) = \phi_c$.
    We then integrate the equations (\ref{D-alpha})-(\ref{D-phi}) outward until $p = 0$, which defines the stellar surface.
    
    \item Physical quantities such as $\phi$, $\phi'$, and $\Lambda'$ must be continuous at the stellar surface. We use this fact and use the values of the variables at $r=R_S$ as initial conditions to solve the equations further for $r>R_S$. 
    
    \item Only one (for shallow effective potentials) or a few symmetric non-trivial values (deep effective potentials) of $\phi(0)$ lead to solutions that decay to $\phi = 0$ at large distances. To determine such values of $\phi_c$, we employ the numerical root-finding routine \texttt{scipy.optimize.root\_scalar}, which iteratively searches for the value of $\phi_c$ that satisfies the required boundary condition at large distances.
    
    \item We repeat this procedure for different choices of $p_c$, obtaining a set of $(M, R_S)$ points, which we then plot to obtain the mass- radius curve. These results are compared with the corresponding mass-radius curve in GR.
\end{enumerate}

\section{Results and Discussions}
We present the results of the numerical calculations and our inferences.
In the following plots, the scalar field mass is expressed in terms of its Compton wavelength
$\lambda_\phi =2\pi/m$.

\subsection{The scalar field profile}
\begin{figure}[tbp]
  \centering
  \includegraphics[width=0.48\linewidth]{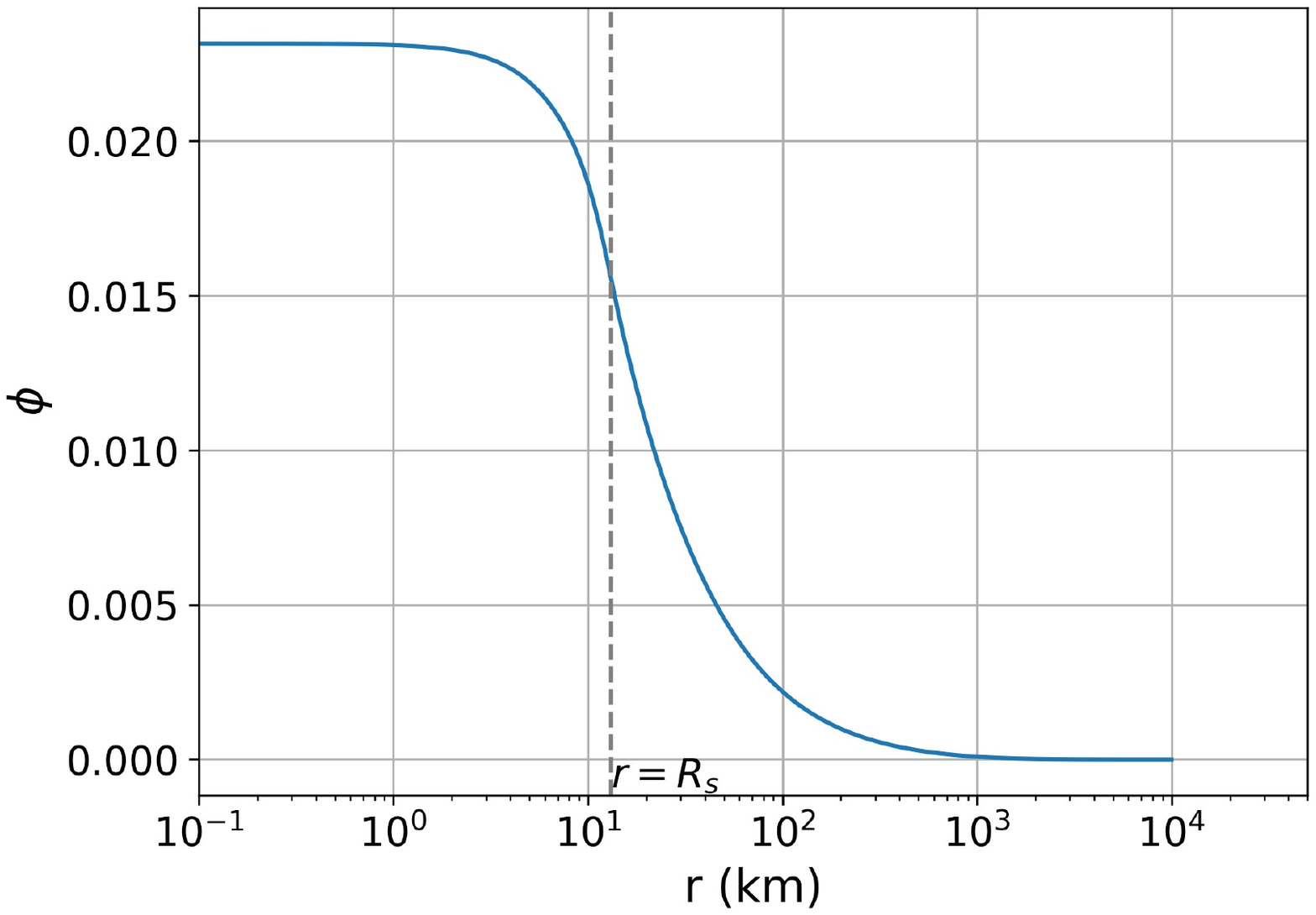}
  \hfill
  \includegraphics[width=0.48\linewidth]{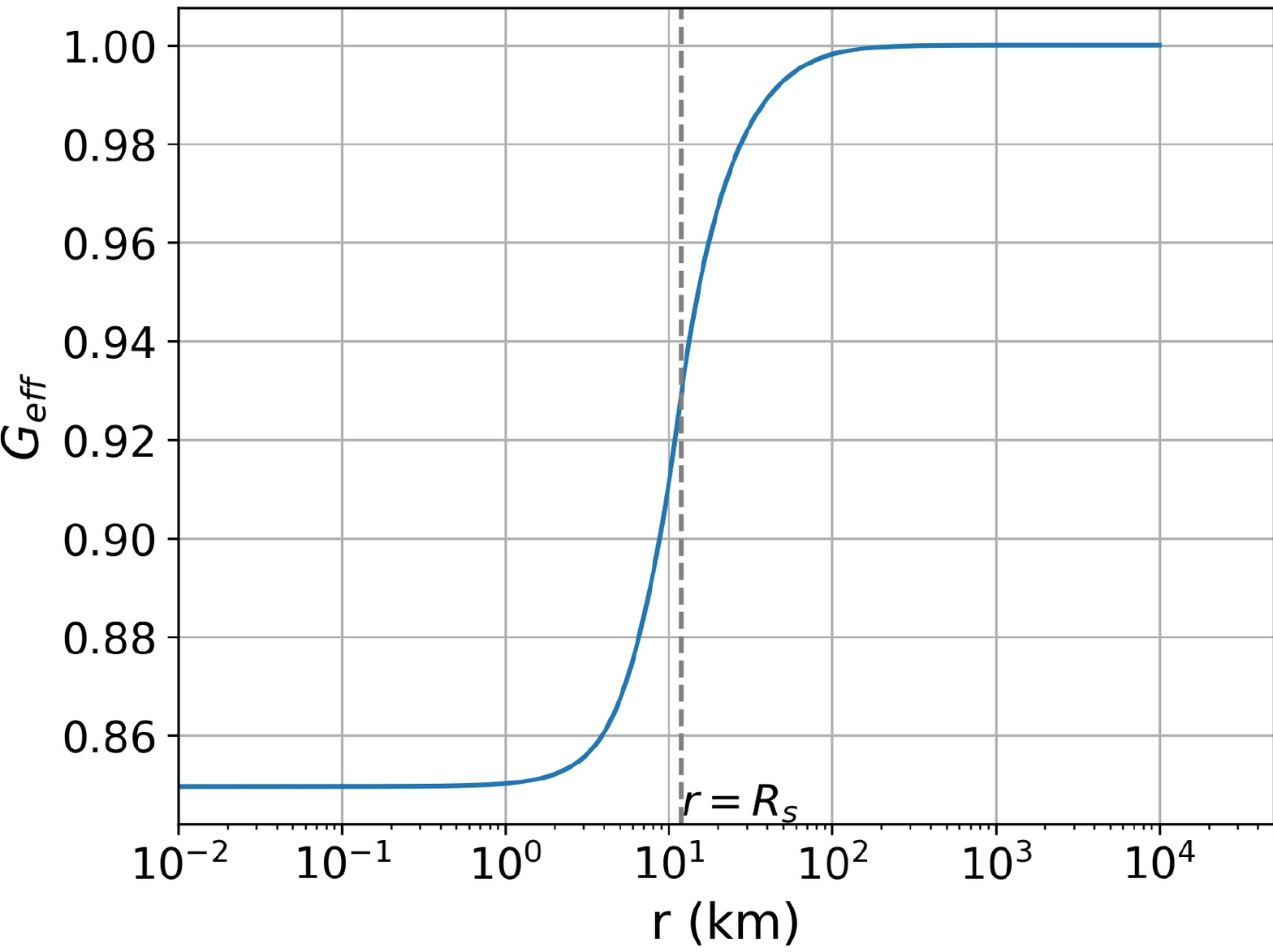}
  \caption{Left panel: scalar field profile for a single non-trivial symmetric solution for $\xi=10$, $\lambda_\phi=6280 \text{ km}$, and $p_c=100\,\text{MeV/fm}^3$.
  Right panel: Effective gravitational constant $G_{\rm eff}$ corresponding to the scalar
  field profile given in the left panel.}
  \label{fig:phi-R_single}
\end{figure}

In this section, we present the scalar field profiles corresponding to various parameter choices for the EoS OPGR(GM1Y4). We begin with a single positive non-trivial solution obtained 
for the parameter values $\lambda_\phi = 6280 \, \text{km}, \xi = 10,  p_c = 100 \, \text{MeV/fm}^3$.
The resultant scalar-field profile is given in the left panel in Fig.~\ref{fig:phi-R_single}. 
As it can be seen, the scalar field starts at a non-zero central value $\phi_c = \phi(r=0)$ 
and decays smoothly to zero outside the star.
The right panel shows the profile of $G_{\rm eff}$ corresponding to the scalar-field profile
in the left panel. 
At the core of the NS, $G_{\rm eff}$ is reduced to $\simeq 0.86$.
This slight reduction of $G_{\rm eff}$ from $G$ is consistent with what we discussed in \ref{Qualitative-behavior}.

\par
\begin{figure}[!htbp]
    \centering
    \includegraphics[scale=0.30]{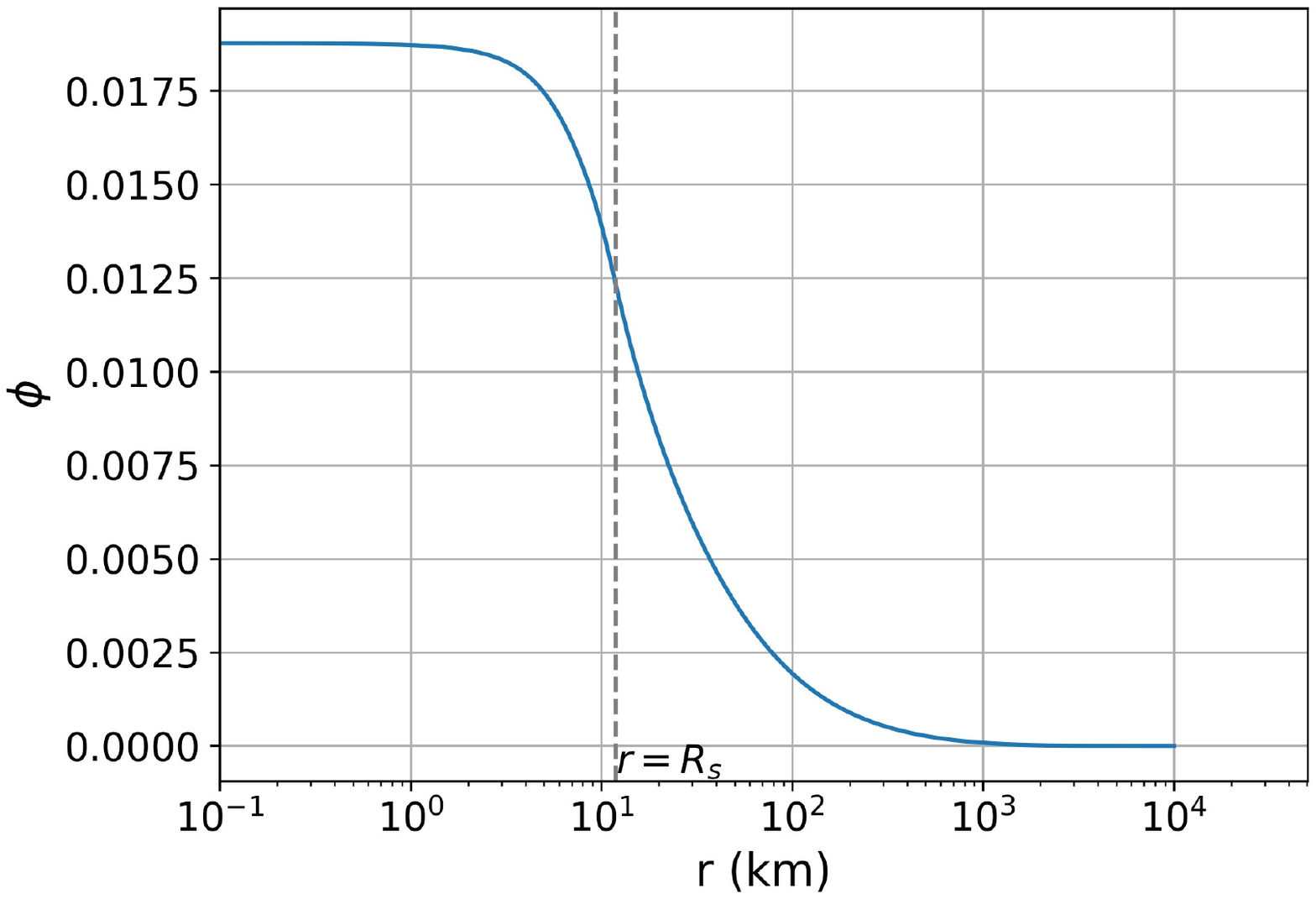}
    \includegraphics[scale=0.30]{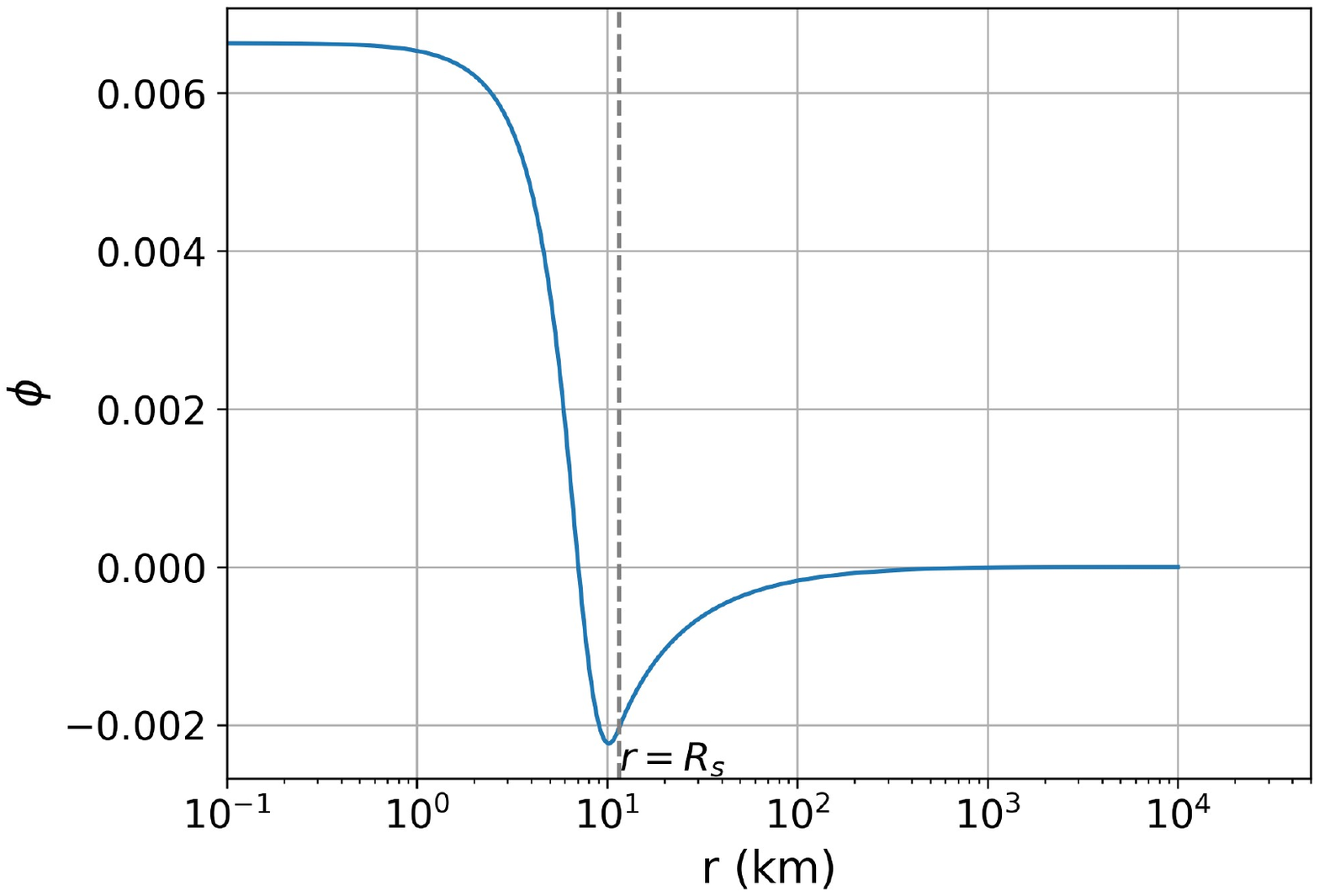}
    \caption{Scalar field profiles for the two non-trivial symmetric solutions obtained for $\xi=20$,$\lambda_\phi=6280 \text{ km}$, and $p_c=300\,\text{ MeV/fm}^3$.}
    \label{fig:phi-R_twosols}
\end{figure}

\vspace{0.35em}
We have confirmed that for larger values of $\xi$ or central pressures, multiple non-trivial symmetric solutions appear. 
For example, Fig.~\ref{fig:phi-R_twosols} illustrates the existence of two distinct solutions for 
$\lambda_\phi=6280 \text{ km},\xi=20,p_c = 300\,\text{ MeV/fm}^3$. 
One of them has a node, i.e.\ the scalar field oscillates once around the minimum before reaching the stellar surface and subsequently decaying to zero.

\begin{figure}[!htbp]
    \centering
    \includegraphics[scale=0.195]{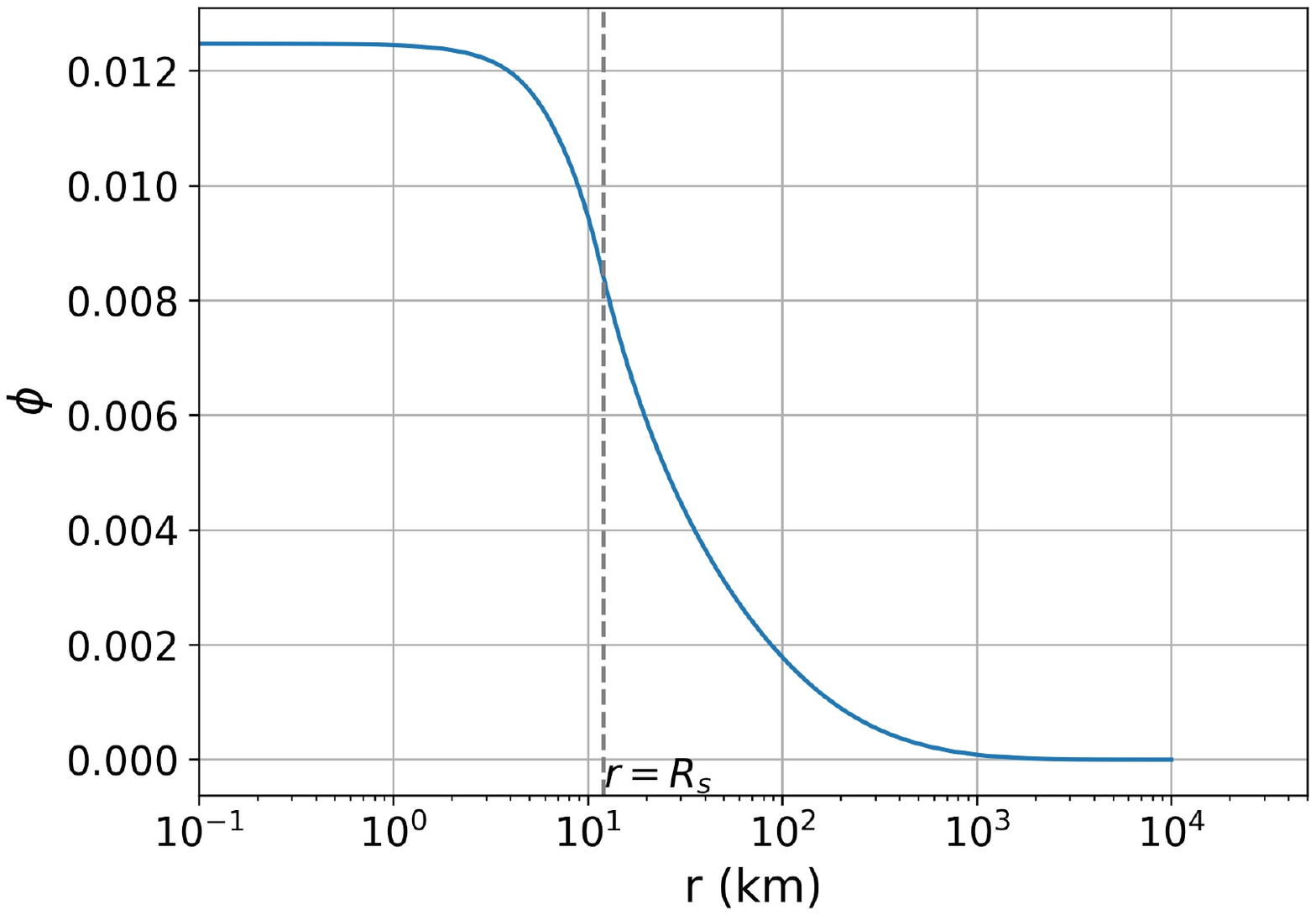}
    \includegraphics[scale=0.195]{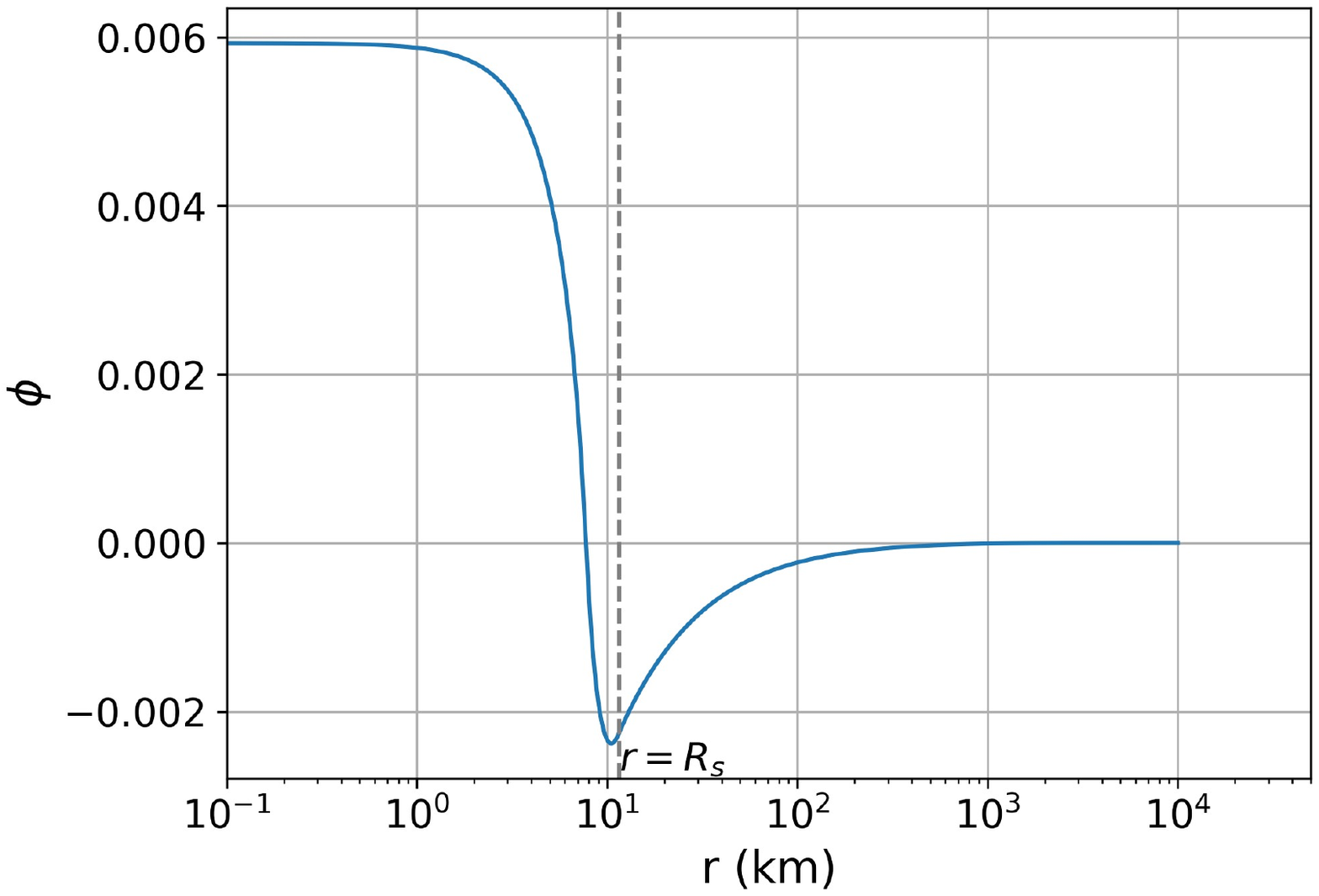}
    \includegraphics[scale=0.195]{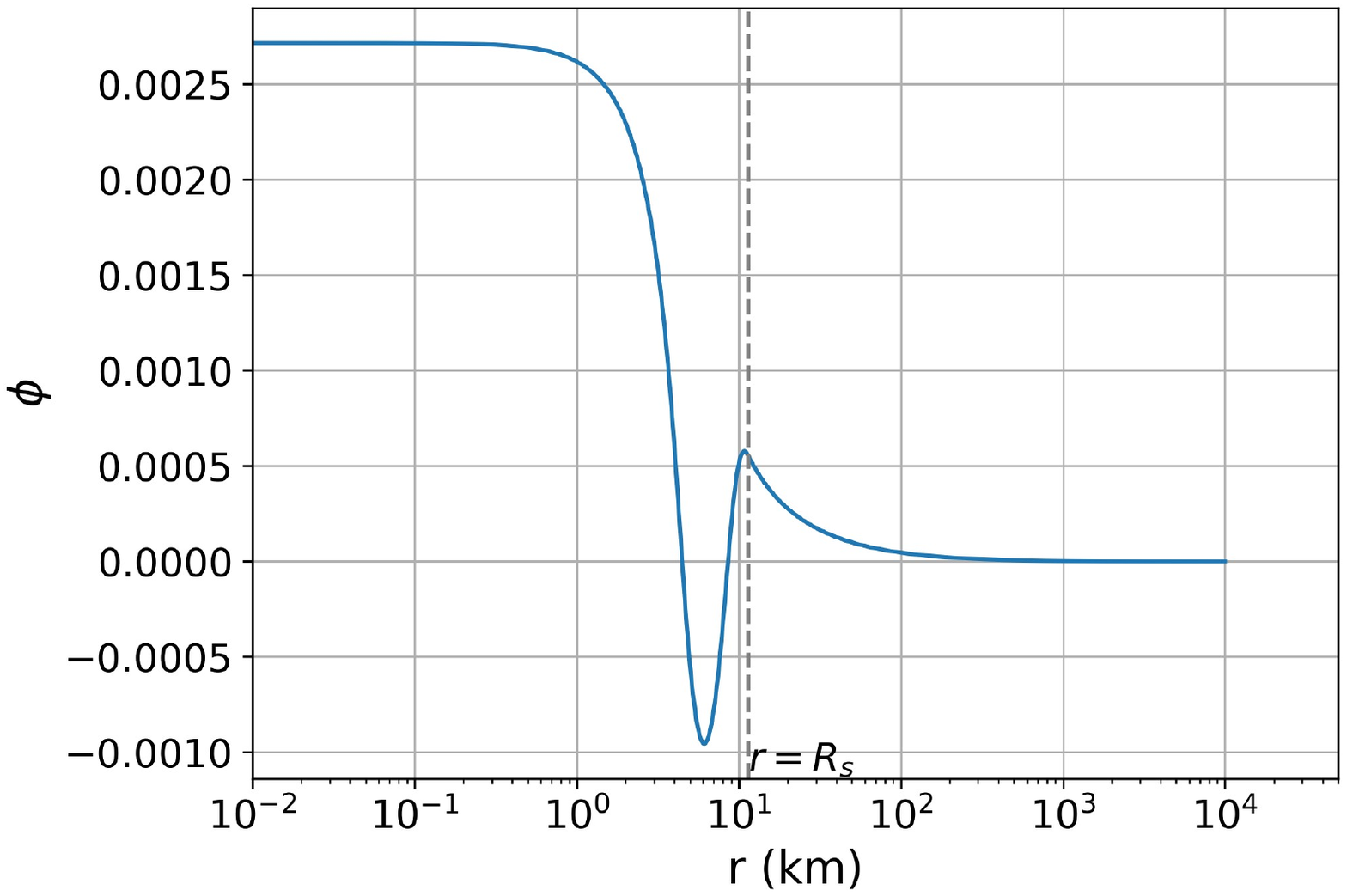}
    \caption{Ground and excited state scalar field profiles for the three non-trivial symmetric solutions obtained for $\xi=50$, $\lambda_\phi=6280 \text{ km}$, and $p_c=300\,\text{MeV/fm}^3$.}
    \label{fig:phi-R_threesols_phi}
\end{figure}

Increasing $\xi$ or $p_c$ further leads to an increasing number of such solutions. In Fig.~\ref{fig:phi-R_threesols_phi}, we show the scalar field profiles for the parameters, 
$\lambda_\phi=6280 \text{ km},\xi=50, p_c=300\,\text{MeV/fm}^3$, for which three symmetric non-trivial solutions exist.
\par\vspace{0.35em}

The nodeless scalar field configuration is generally expected to be dynamically stable, whereas solutions with nodes represent excited states that are typically unstable (This has also been confirmed in the dynamical analysis done in \cite{Degollado2024}). We also find that the mass–radius curves obtained for nodeless solutions yield a higher maximum mass compared to those with nodes. Therefore, in the following analysis, we focus on the nodeless scalar-field configurations when constructing the mass–radius curves.

\FloatBarrier

\subsection{The dependence of Mass-Radius curve on \texorpdfstring{$\xi$}{xi} and \texorpdfstring{$m$}{m}}

The main motivation of our work is to demonstrate that NSs with maximum masses $M_{\text{max}} \geq 2 M_\odot$ can exist even when their interiors contain hyperons or other forms of strange matter. 
To this end, we present the resulting mass--radius (M--R) relations for such EoS, 
exploring variations in the coupling constant $\xi$, which controls the strength of the scalar--gravity interaction, and the scalar-field mass $m$.

\subsubsection{Dependence of the Mass-Radius Curve on \texorpdfstring{$m$}{m}}

We begin by examining how the mass of the scalar field $m$ affects the mass--radius (M--R) relation. Throughout this analysis, we fix the coupling constant at $\xi = 10$. 
\par\vspace{0.35em}
We computed Mass-Radius curves for three different values of $\lambda_\phi$: 
$\lambda_\phi \sim 628,\; 6280,\; 62800\ \text{km}$.
For these values, as the analytic expression for the critical density given by 
Eq.~(\ref{critical-density}) indicates, 
the spontaneous scalarization occurs inside the NS.
These values of $\lambda_\phi$ are compatible with observational constraints, which place an upper limit of $\lambda_\phi \leq10^7$ km (This bound follows from the orbital scale of the binary pulsar PSR J0348–0432 \cite{Antoniadis2013}; for scalar Compton wavelengths smaller than this scale, the scalar field would not significantly affect the orbital dynamics.).
The resulting mass-radius curves for the OPGR(GM1Y4) and BHB(DD2$\Lambda$) EoS are shown in Fig.~\ref{fig:mass_rad_xi=10,gamma=0_varmass_OPGR(GM1Y4)} 
and Fig.~\ref{fig:mass_rad_xi=10,gamma=0.1_varmass_BHB(DD2L)}, respectively.

\begin{figure}[!htb]
    \centering
    \includegraphics[scale=0.30]{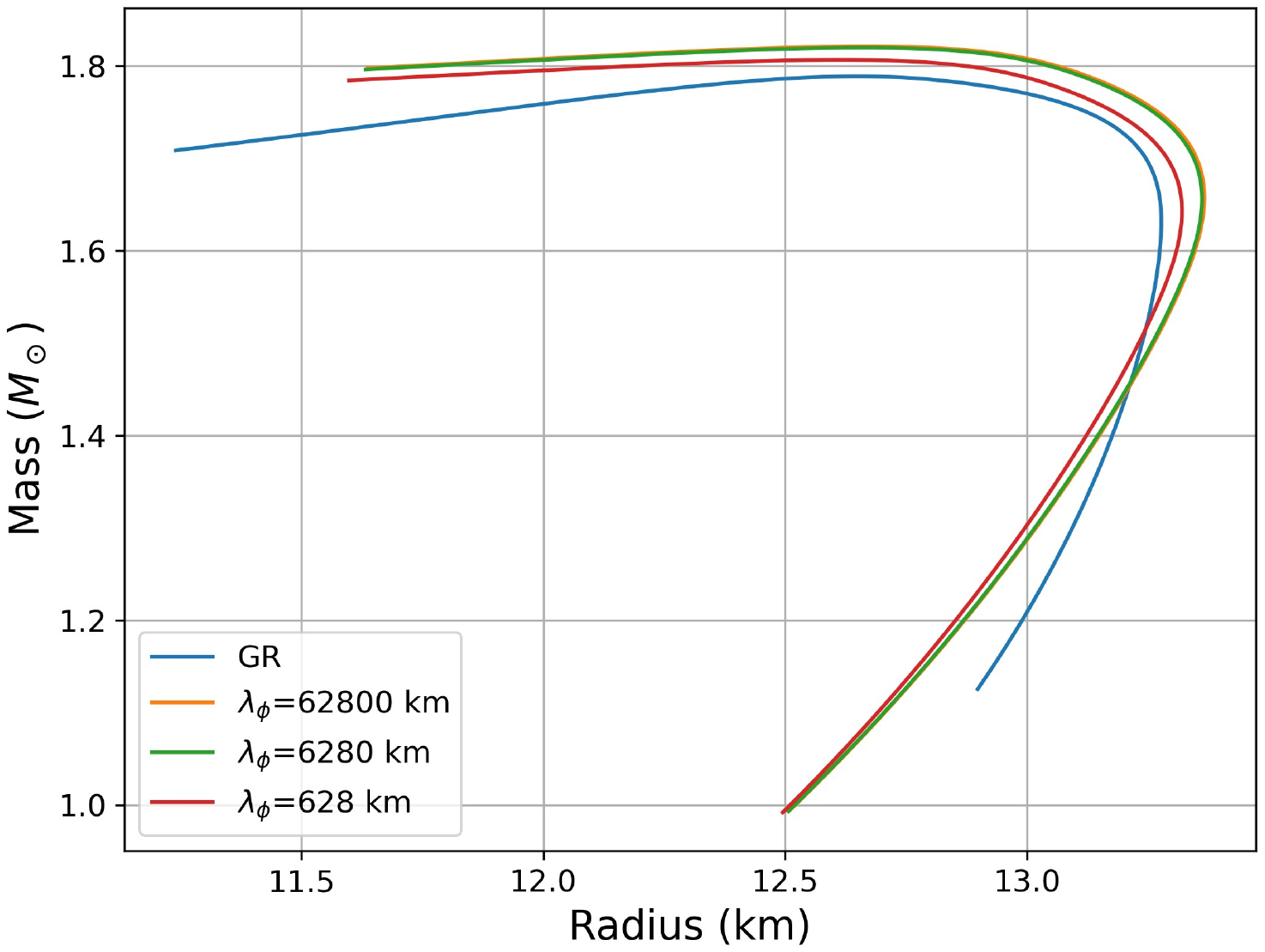}
    \includegraphics[scale=0.30]{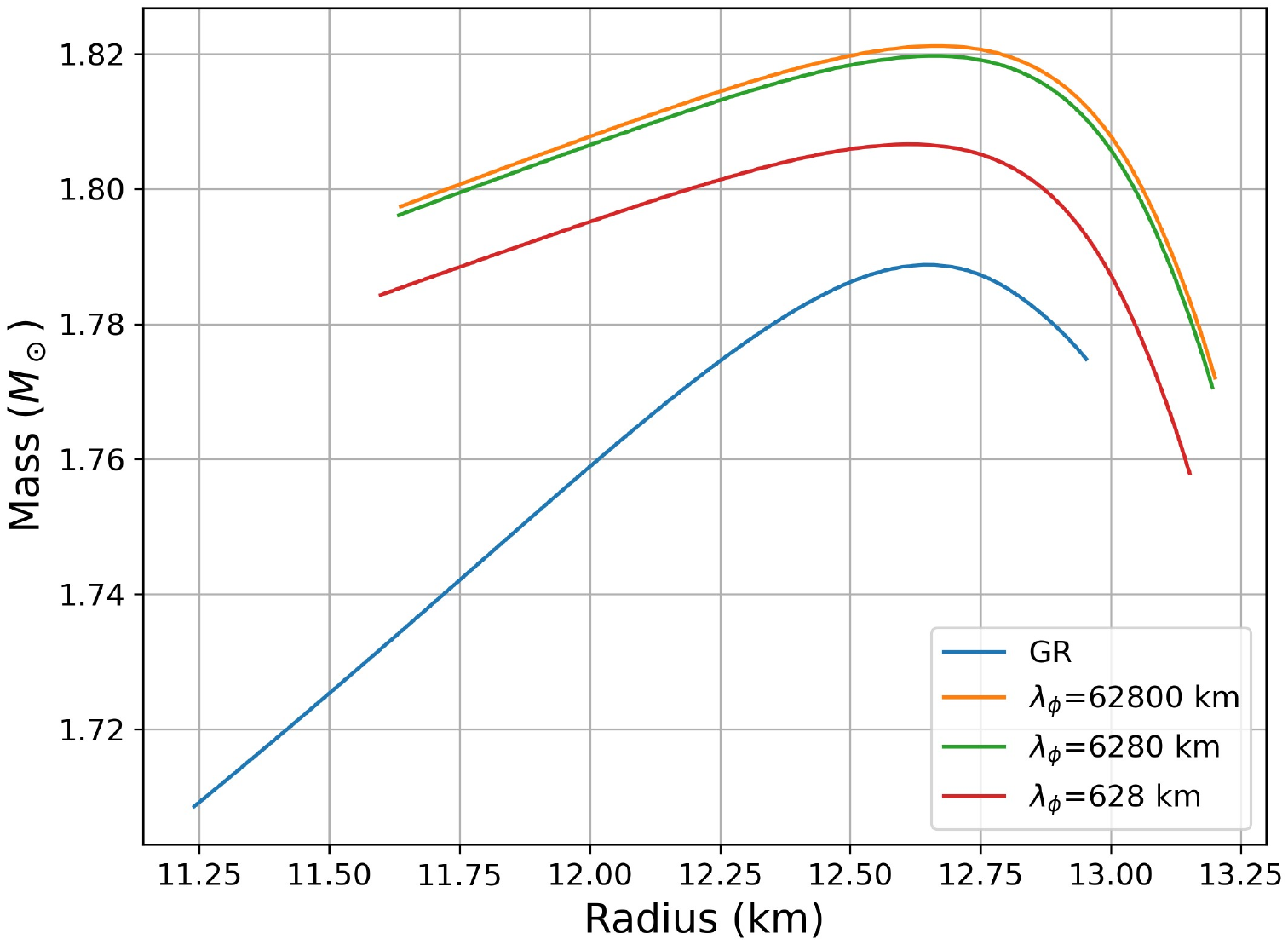}
    \caption{Mass--radius curves for OPGR(GM1Y4) for varying scalar field masses (left panel). 
    The right panel shows a magnified view around the region containing $M_{\rm max}$.}
    \label{fig:mass_rad_xi=10,gamma=0_varmass_OPGR(GM1Y4)}
\end{figure}

\begin{figure}[!htbp]
    \centering
    \includegraphics[scale=0.3]{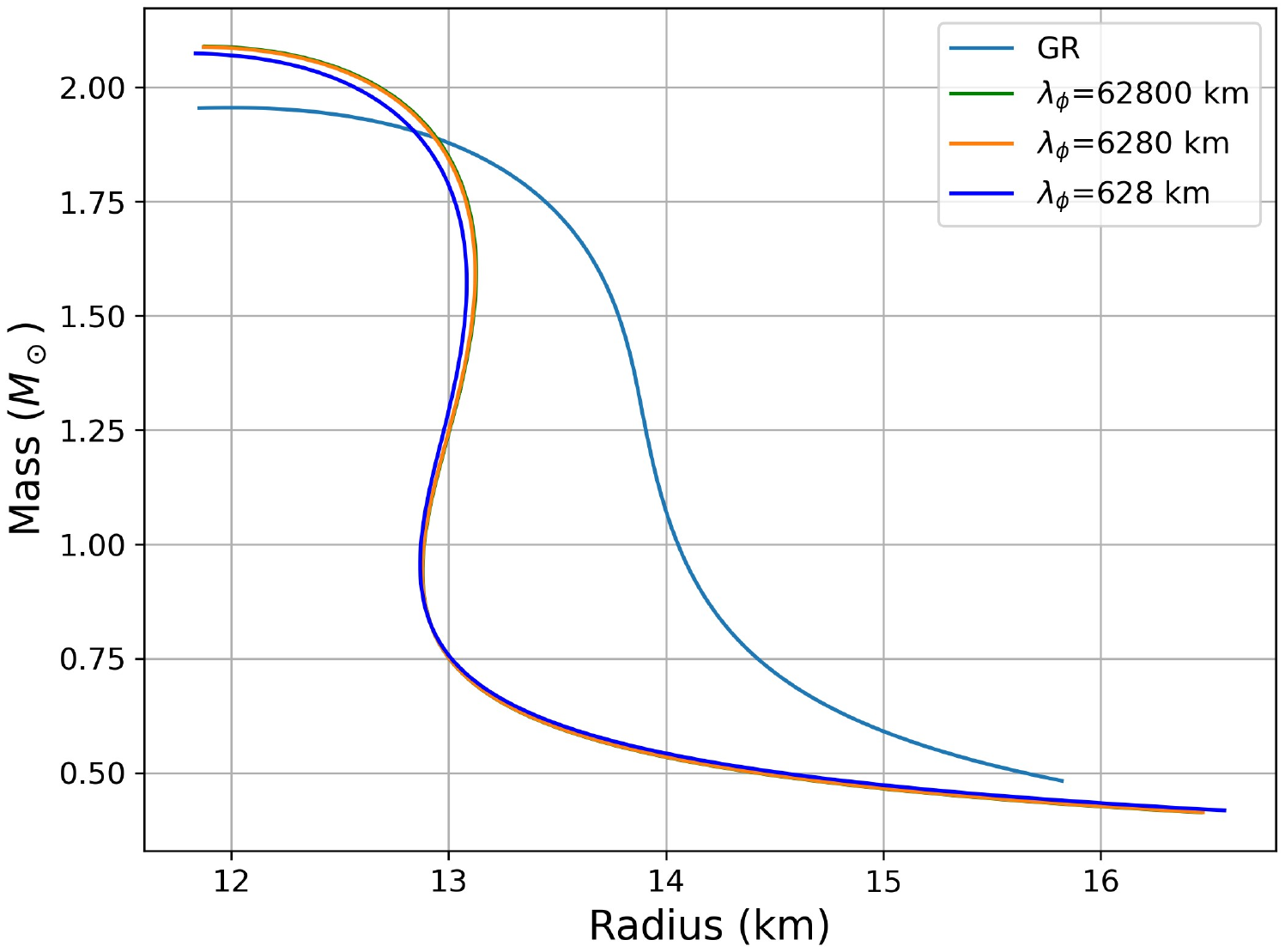}
    \includegraphics[scale=0.3]{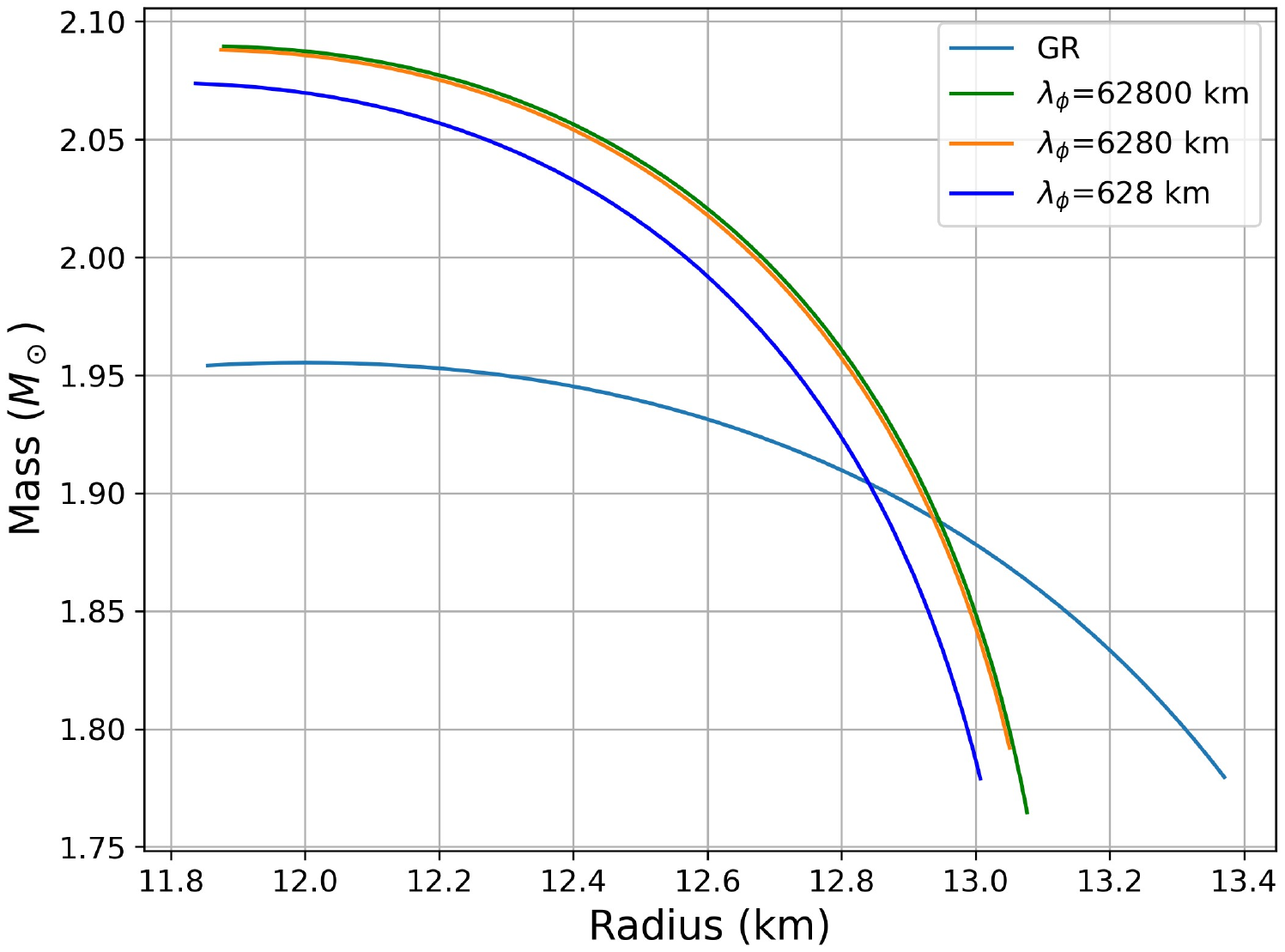}
    \caption{Mass--radius curves for BHB(DD2$\Lambda$) for varying scalar field masses (left panel). 
    The right panel shows a magnified view around the region containing $M_{\rm max}$.}
    \label{fig:mass_rad_xi=10,gamma=0.1_varmass_BHB(DD2L)}
\end{figure}

\par\vspace{0.35em}
We observe that the maximum mass of the neutron star, $M_{\rm max}$, increases compared to its value in GR.
This is consistent with the intuitive view that the weakening of gravity inside the NS allows
the nuclear matter pressure to support a heavier NS that would otherwise be unstable in GR. 
The increase in $M_{\rm max}$ depends on the nuclear EoS; however, the relative
change in $M_{\rm max}$ is not significant in any case, remaining at most approximately $6\%$.
This is a natural consequence of the fact that the deviation of $G_{\rm eff}$ 
from $G$ is of the same order.
The mass-radius curves in both Fig.~\ref{fig:mass_rad_xi=10,gamma=0_varmass_OPGR(GM1Y4)} 
and Fig.~\ref{fig:mass_rad_xi=10,gamma=0.1_varmass_BHB(DD2L)} become insensitive to 
$\lambda_\phi$ in the large Compton wavelength limit,
which is consistent with the qualitative argument made at the end of \ref{Qualitative-behavior}.
\FloatBarrier
\subsubsection{Dependence of the Mass Radius Curve on \texorpdfstring{$\xi$}{xi}}\label{variation_xi}
Next, we examine how the coupling constant $\xi$, associated with the Ricci scalar, influences the mass--radius relation. 
We carry out this analysis for $\xi =10,\; 20,\; 50$
while keeping the scalar mass fixed at $\lambda_\phi=6280 \text{ km}$. 
The resulting Mass-Radius curves for the OPGR(GM1Y4) and BHB(DD2$\Lambda$) EoS are shown in Fig.~\ref{fig:mass_rad_m=1e-3,gamma=0.1_varxi_OPGR(GM1Y4)} and ~\ref{fig:mass_rad_m=1e-3,gamma=0.1_varxi_BHB(DD2L)}, respectively.

\begin{figure}[!htbp]
    \centering
    \includegraphics[scale=0.3]{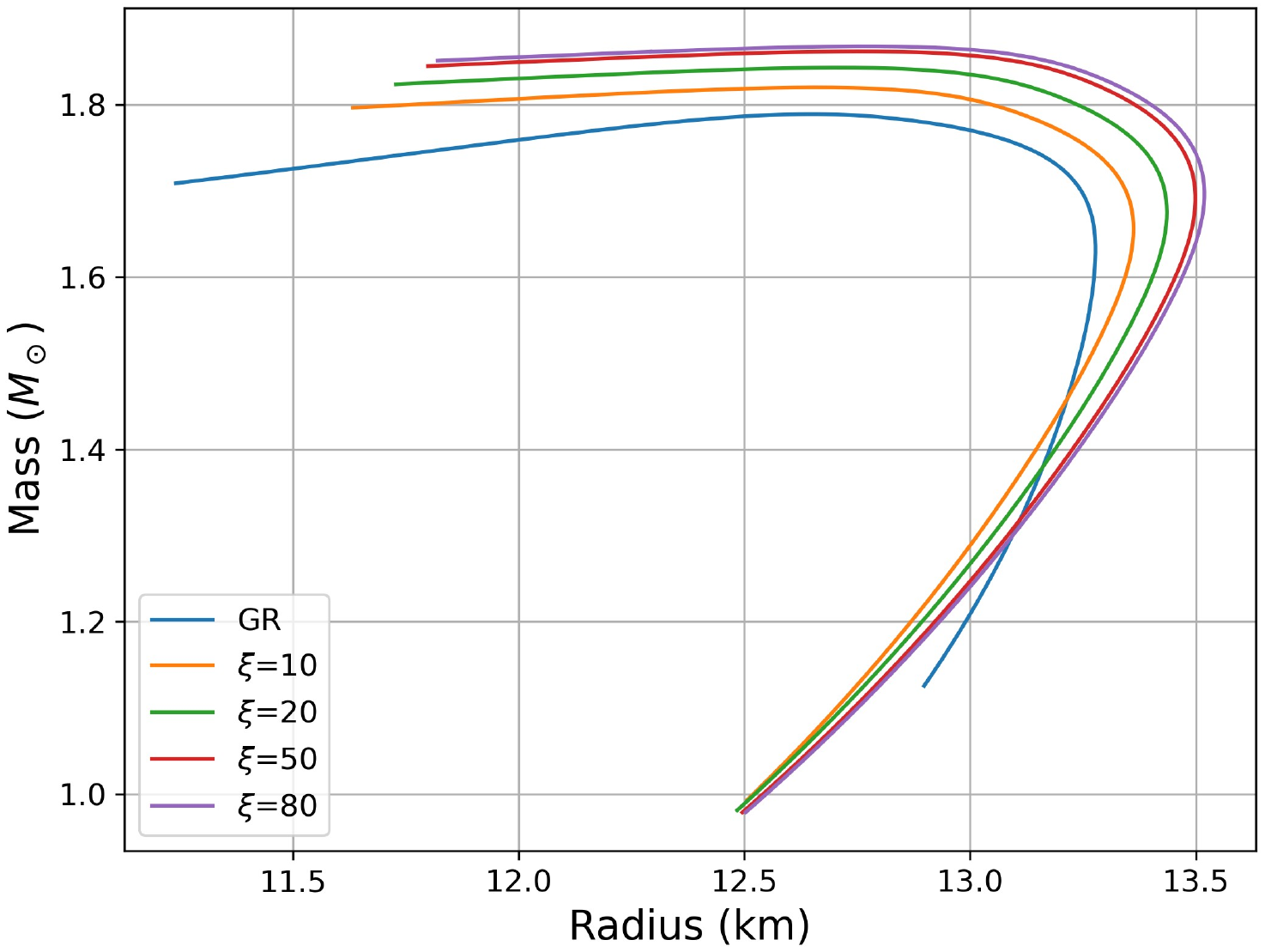}
    \includegraphics[scale=0.3]{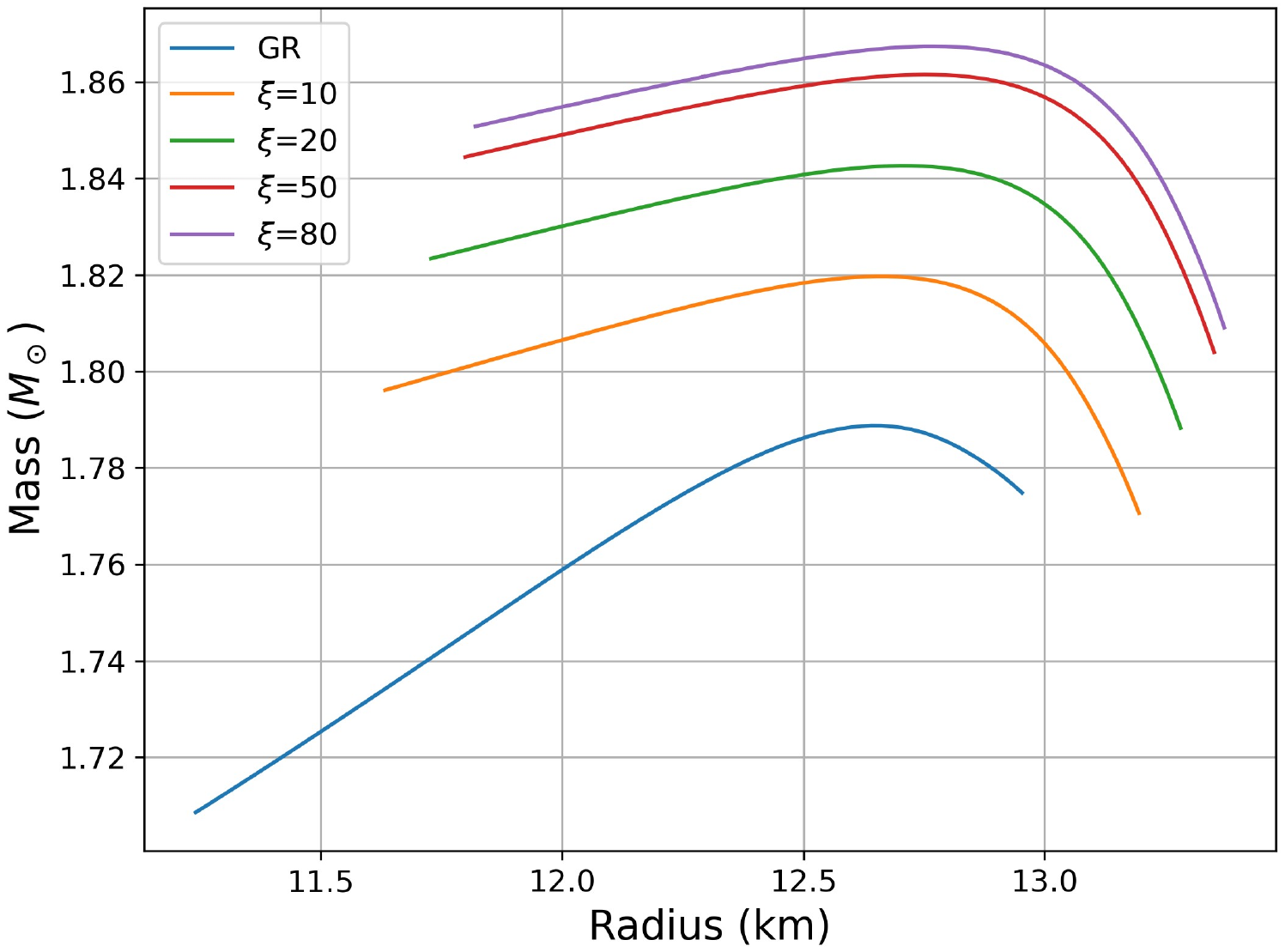}
    \caption{Mass--radius curves for OPGR(GM1Y4) for varying values of the coupling constant $\xi$ (left panel). The right panel shows a magnified view of the region containing $M_{\rm max}$.}
    \label{fig:mass_rad_m=1e-3,gamma=0.1_varxi_OPGR(GM1Y4)}
\end{figure}

As illustrated in the plots, the maximum mass $M_{\rm max}$ increases monotonically with $\xi$. 
Since $\xi$ governs the strength of spontaneous scalarization, this trend is a natural consequence 
of the increased coupling.
To examine this behavior more closely, we again consider the effective potential $V_{\rm eff}$ 
defined in Eq.~(\ref{def-effective-potential}). 
In the large-$\xi$ limit, its derivative reduces to 

\begin{equation}
\frac{dV_{\rm eff}}{d\phi} = -\frac{\rho + m^2 \phi^2}{\phi}.
\end{equation}

\par\vspace{0.35em}
Notably, this expression becomes independent of $\xi$, rendering the equation of motion for $\phi$ insensitive to further increases in $\xi$. 
This suggests that the effective gravitational constant scales roughly as 
$G_{\rm eff} \propto 1/\xi$ in this limit.
\par\vspace{0.35em}
Conversely, the numerators on the right-hand side of the gravitational field equations 
(\ref{einstein-eq}) contain terms proportional to $\xi$. 
In the large-$\xi$ limit, where these terms dominate, the $\xi$ in the numerator and the $1/\xi$ scaling from the scalar field effectively cancel, 
making the right-hand side of Eq.~(\ref{einstein-eq}) nearly independent of $\xi$. 
This qualitative argument explains the behavior observed in Figs.~\ref{fig:mass_rad_m=1e-3,gamma=0.1_varxi_OPGR(GM1Y4)} and \ref{fig:mass_rad_m=1e-3,gamma=0.1_varxi_BHB(DD2L)}: the variation in $M_{\rm max}$ is relatively modest ($\lesssim 10\%$) compared to the order-of-magnitude variations in $\xi$.

\begin{figure}[!htbp]
    \centering
    \includegraphics[scale=0.3]{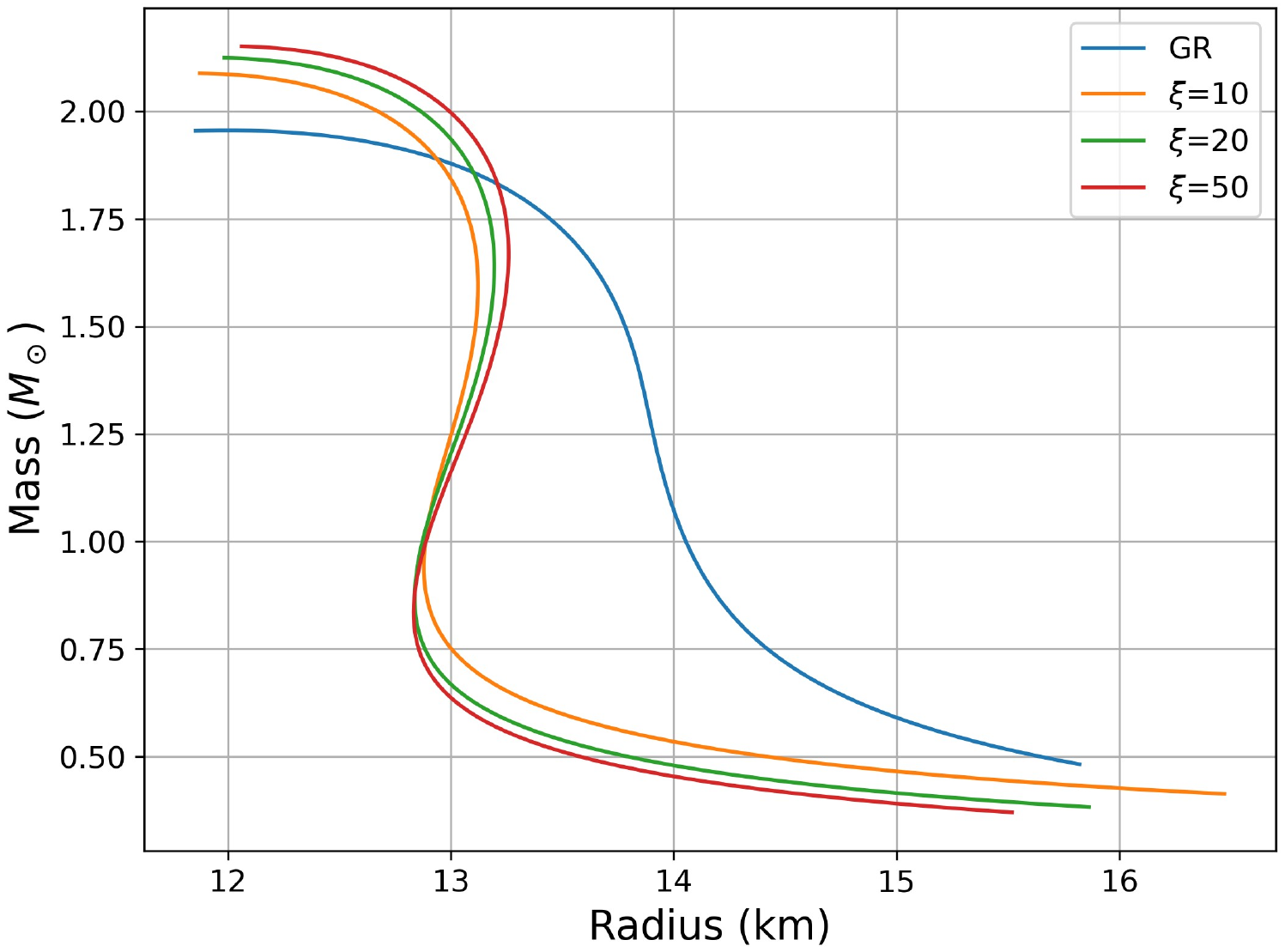}
    \includegraphics[scale=0.3]{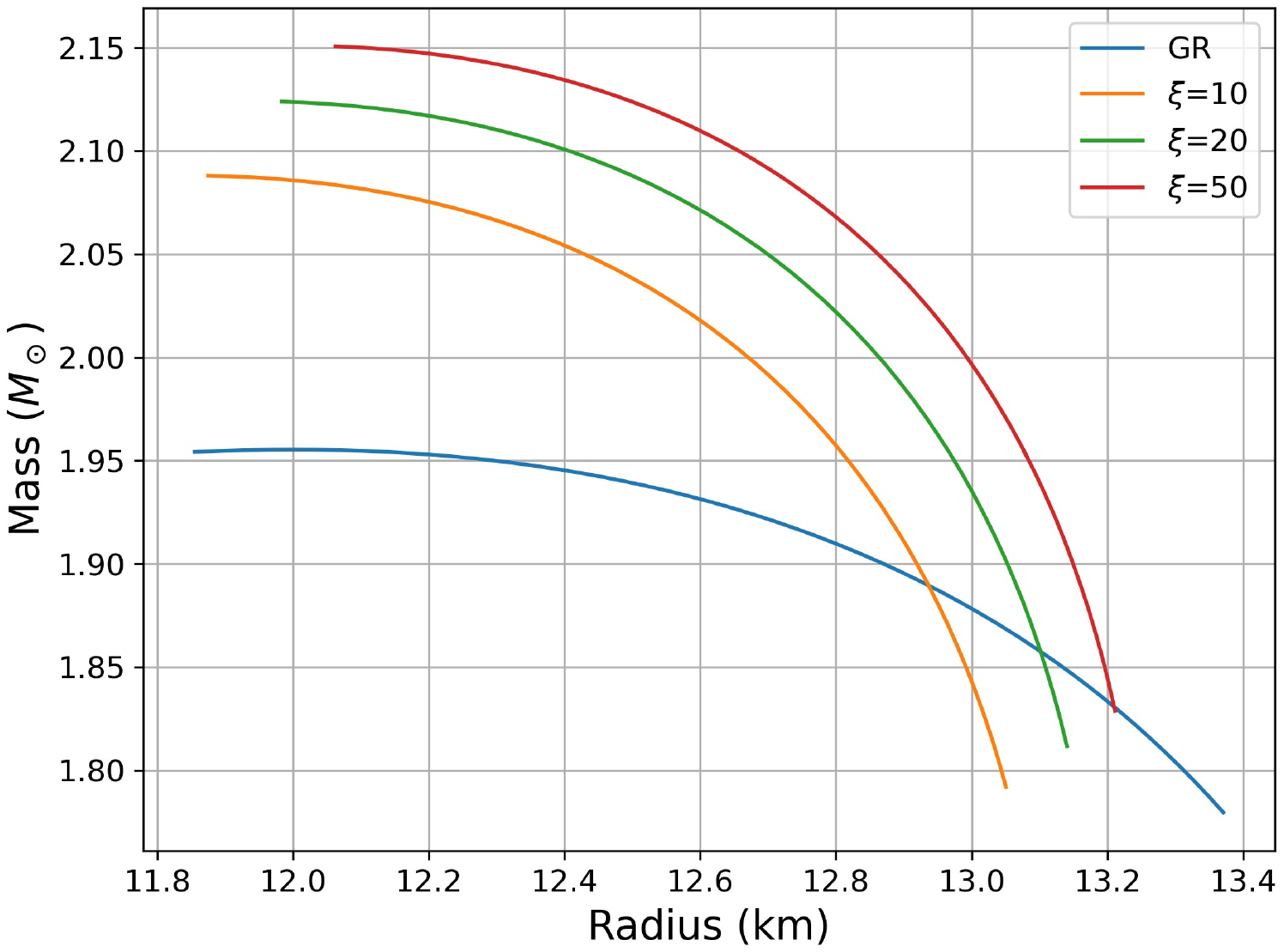}
    \caption{Mass--radius curves for BHB(DD2$\Lambda$) for varying values of the coupling constant $\xi$ (left panel). The right panel shows a magnified view of the region containing $M_{\rm max}$.}
    \label{fig:mass_rad_m=1e-3,gamma=0.1_varxi_BHB(DD2L)}
\end{figure}
\subsection{Effects of self-interaction}

We have so far assumed that the scalar field $\phi$ is a free field except that it has 
coupling to the Ricci scalar.
In this section, we explore the consequences of including a self-interaction term
$\frac{\gamma}{4}\phi^4 $ in the action of the theory.
The action is given by
\begin{equation}
S=\int d^4x\sqrt{-g}\Bigl(\frac{R}{16\pi }-\frac{(\nabla\phi)^2}{2}-\frac{m^2\phi^2}{2}+\frac{\xi R\phi^2}{2}-\frac{\gamma \phi^4}{4}\Bigr)+ S_M[A,g_{\mu\nu}]. \label{action2}
\end{equation}
We solved the equations (\ref{D-alpha})-(\ref{D-phi}) with the self-interaction term
being included under the same boundary conditions.
The Mass-Radius curves for different values of $\gamma$ are presented in Fig.~\ref{fig:mass_rad_m=1e-3,xi=10_vargamma_OPGR(GM1Y4)}.

\begin{figure}[!htbp]
    \centering
        \includegraphics[scale=0.3]{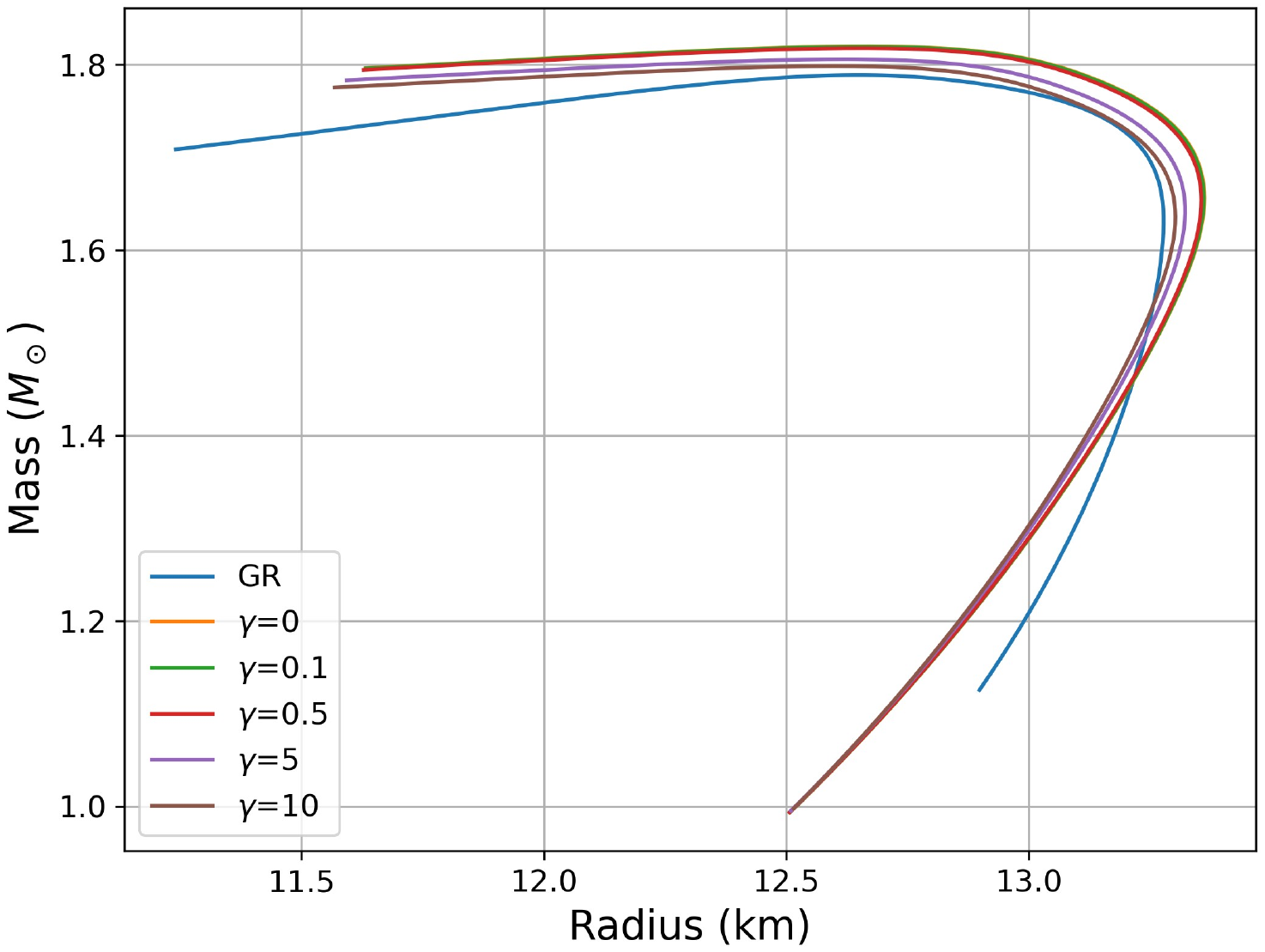}
        \includegraphics[scale=0.3]{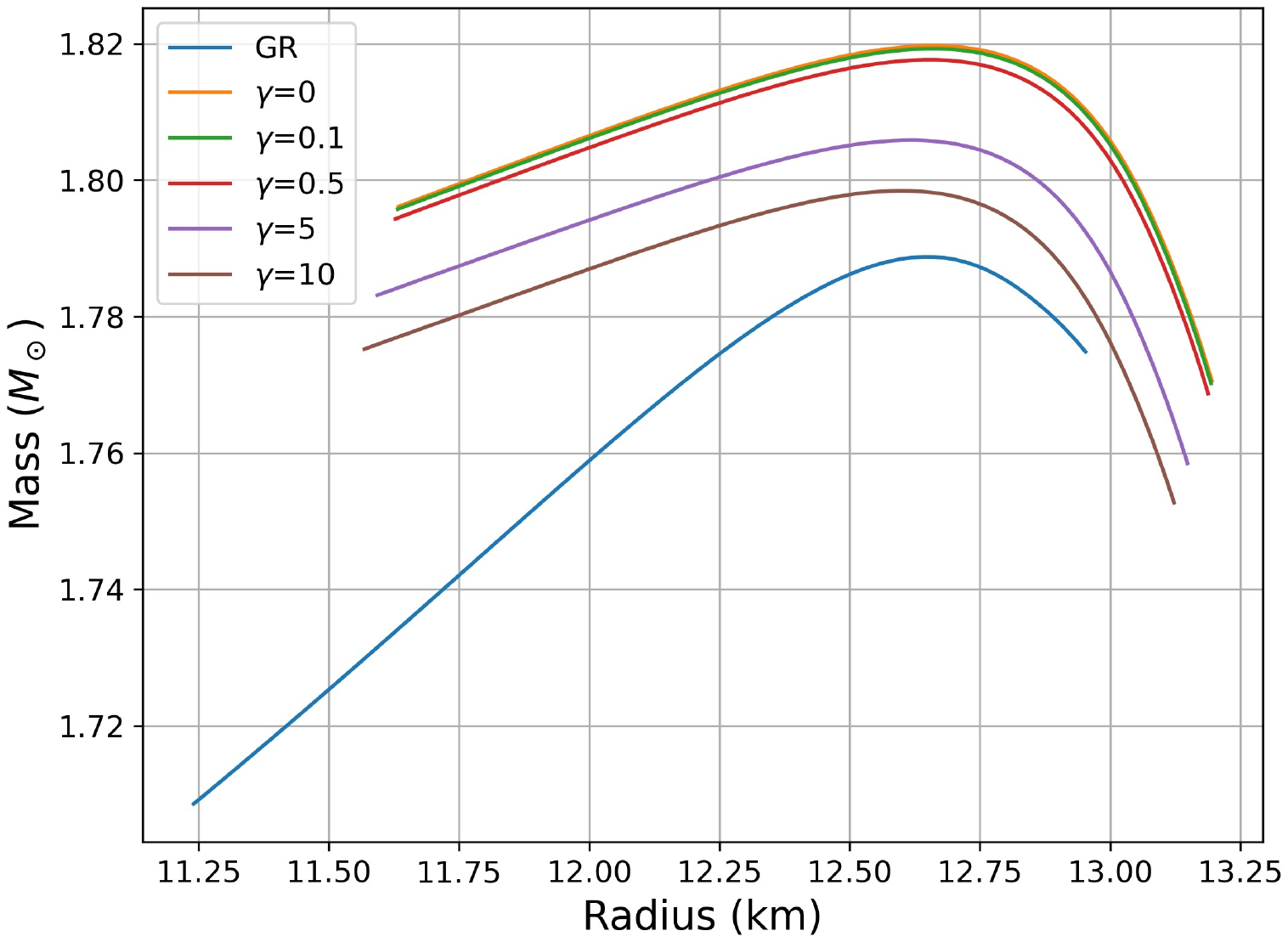}
    \caption{Mass--radius curve for OPGR(GM1Y4) for varying values of the coupling constant $\gamma$. }
    \label{fig:mass_rad_m=1e-3,xi=10_vargamma_OPGR(GM1Y4)}
\end{figure}

\begin{figure}[!htbp]
        \centering
        \includegraphics[scale=0.44]{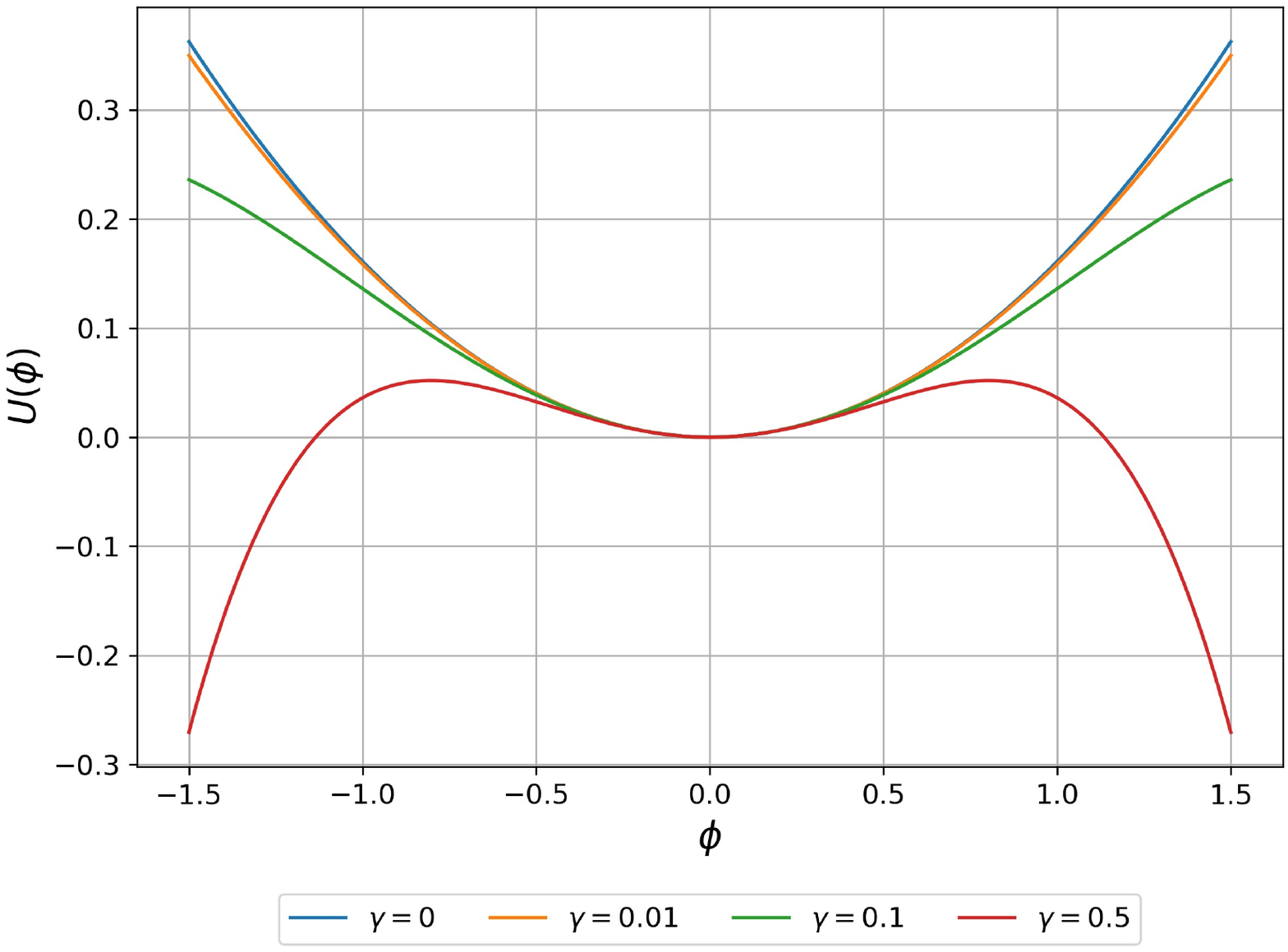}
    \caption{Effective potential $U(\phi) = -V_{\text{eff}}(\phi)$ for different parameters.}
    \label{fig:varpotential_xi_gamma}
\end{figure}

We find that the self-interaction term acts to suppress scalarization, and increasing values of $\gamma$ lead to a reduction of the maximum mass of the neutron star. 
The underlying reason for this behavior is related to the shape of the potential $U=-V_{\rm eff}$
introduced in \ref{Qualitative-behavior}.
The plot $U(\phi)$ in the present case for several values of $\gamma$ is given in Fig.~\ref{fig:varpotential_xi_gamma}.
We observe that the potential height decreases as $\gamma$ increases.
Thus, as the value of $\gamma$ is increased, 
the corresponding initial value of $x$ (i.e., the value of $\phi$ at the center of the NS)
satisfying the boundary condition becomes closer to the origin,
rendering the degrees of the spontaneous scalarization weaker.

\subsection{Consequence on the hyperon puzzle}

Historically, the appearance of hyperons in the neutron-star core was found to significantly soften the equation of state, typically reducing the maximum mass of neutron stars to values around $\sim1.4\,M_\odot$. More recent studies indicate that additional repulsive hyperon interactions can lead to considerably stiffer hyperonic matter \cite{Weissenborn2012}. The hyperonic EoS considered here, OPGR(GM1Y4) and BHB(DD2$\Lambda$), yield maximum masses of $M_{\rm max}\approx1.79\,M_\odot$ and $M_{\rm max}\approx1.95\,M_\odot$, respectively, within general relativity.
\par\vspace{0.35em}
The present model provides a mechanism to enhance the maximum mass of neutron stars without introducing additional couplings in the matter sector, instead relying on a non-minimal coupling induced by the Ricci scalar. However, for the EOS OPGR(GM1Y4), the enhancement is not sufficiently large to account for neutron stars with masses $\sim 2M_\odot$. Among the two well-known $\sim 2M_\odot$ measurements, we focus primarily on the Shapiro-delay mass of PSR~J1614--2230, $M = 1.97\,M_\odot$, since the mass computed numerically in this work corresponds to the gravitational mass. In our framework, the scalar field extends beyond the stellar surface, leading to a non-trivial contribution to the mass distribution outside the star.
\par\vspace{0.35em}
For OPGR(GM1Y4), the maximum mass supported within GR is approximately $1.79\,M_\odot$. In the modified theory, the maximum mass increases for positive values of $\xi$, and grows monotonically with increasing $\xi$. The largest mass obtained in this case is about $1.865\,M_\odot$ for $\xi = 80$, corresponding to an enhancement of roughly $4\%$. This remains below the observational benchmark of $1.97\,M_\odot$. The relatively modest increase can be attributed to the behavior of the effective gravitational coupling $G_{\rm eff}$, which decreases significantly in this regime. As discussed earlier in \ref{variation_xi}, for sufficiently large $\xi$, $G_{\rm eff}$ scales approximately as $1/\xi$, and the enhancement from scalarization is partially offset by this reduction in the effective gravitational strength. Consequently, even in the modified theory, this EOS remains incompatible with the $\sim 2M_\odot$ constraint.
\par\vspace{0.35em}
In contrast, for BHB(DD2$\Lambda$), the maximum mass in GR is already about $1.95\,M_\odot$. Within the modified theory, this increases to approximately $2.15\,M_\odot$ for $\xi = 50$ and $\lambda_\phi = 6280\,\mathrm{km}$, corresponding to an enhancement of nearly $10\%$. This brings the model comfortably above the $2M_\odot$ threshold. Although this EOS already supports a relatively large maximum mass even with hyperons included, the additional enhancement from scalarization makes it fully compatible with current observational constraints. This demonstrates that the degree of mass enhancement is not universal, but depends sensitively on the underlying equation of state.

\section{Conclusions}

This work investigated dark matter modeled as a scalar field non-minimally coupled to gravity that can trigger spontaneous scalarization in neutron stars through tachyonic instabilities, and how this mechanism modifies their mass–radius structure. The primary aim was to investigate whether such a framework can produce stars with masses $\geq 2\, M_\odot$ even when hyperonic models, known to soften the equation of state, are included.
\par\vspace{0.35em}
We obtained the mass--radius relations for neutron stars by deriving the equations of motion from the action and solving the resulting four coupled differential equations. The behaviour of the mass–radius curve was examined for different combinations of the parameters $\lambda_\phi, \text{ and } \xi,$\, while imposing observational constraints from PSR J0348--0432 on the scalar Compton wavelength.
\par\vspace{0.35em}
Our analysis shows that scalarization caused by the inclusion of dark matter can increase the maximum mass of neutron stars beyond $2M_\odot$ for certain equations of state, including hyperonic ones. However, although the reduced effective gravitational constant weakens gravity and permits more matter to accumulate before collapse, the strong suppression of gravity inside the neutron star also reduces the resulting gravitational mass, limiting the overall mass enhancement. Nonetheless, this study demonstrates that the maximum mass of neutron stars can be raised well above the general relativistic predictions in all cases. We also identified multiple non-trivial symmetric solutions at higher coupling strengths and attributed these to enhanced scalarization. The corresponding scalar-field profiles displayed oscillatory behavior inside the star before decaying outside, highlighting the non-trivial scalar-field structure that emerges in this regime.
\par\vspace{0.35em}
The modifications to neutron-star structure induced by tachyonic instabilities and scalarization may leave observable imprints on gravitational-wave signals. With the increased sensitivity of upcoming third-generation detectors, parameters such as tidal deformability will be measured with unprecedented precision \cite{Punturo2010}. As tidal deformability is directly linked to the equation of state and the equations governing neutron-star structure \cite{Hinderer2008, Creci2023}, these measurements could provide indirect evidence for additional components affecting the stellar interior, such as dark matter coupled to gravity. Such observations may therefore offer insight into the role of non-minimally coupled dark matter in neutron-star interiors, where its gravitational interaction could modify stellar structure and potentially provide a new perspective on the hyperon puzzle.

\section*{Acknowledgement}
T.S. gratefully acknowledges support from JSPS KAKENHI grant (Grant Number JP23K03411).

\begin{appendices}
    \section{Explicit expressions of \texorpdfstring{$A_0, A_1, B_1, B_2, C_1, C_2$}{A0, A1, B1, B2, C1, C2}} 
    \label{appendix-coefficients}
\[A_1=1+\eta+8\pi r^2p-4\pi r^2m^2\phi^2\]
\[A_0=  -1-\eta +4 \pi  r^2\phi'^2-32\pi\xi r\phi\phi'\]

\[
    \begin{split}
    B_1 &= -(\eta +1)^3 + 64 \pi^2  \xi^2 \phi' r^3 \phi\eta \left(2 m^2 \phi^2 + 4 \rho \right) + 32 \pi^2 (\eta +1)  \xi \phi' r^3 \phi (m^2 (4 \xi +2) \phi^2 + 2 \rho )  \\
    &\quad - 4 \pi (\eta +1)  \xi r^2 \eta \left( 2 m^2 \phi^2 - 8 \rho \right)- 64 \pi^2 (\eta +1)  \xi p r^2 \phi (\phi' r - 6 \xi \phi)- 96 \pi (\eta +1)  \xi^2 \phi' r \phi\eta \\
    &\quad   - 6 \xi\eta (\eta +1)^2+ 2 \pi (\eta +1)^2  r^2 \left(m^2 (8 \xi +2) \phi^2 + 4 \rho \right)- 16 \pi (\eta +1)^2  \xi \phi' r \phi 
    \end{split}
\]

\[
    \begin{split}
    B_0&=(\eta +1)^3-128 \pi ^2  \xi ^2 \phi'^3 r^3 \phi\eta+192 \pi  \xi ^2 \phi'^2 r^2 \eta^2+128\pi ^2 (\eta +1)  \xi ^2 \phi'^3 r^3 \phi \\
    &\quad +40 \pi (\eta +1)  \xi \phi'^2 r^2 \eta+96\pi (\eta +1)  \xi ^2 \phi' r \phi\eta+6  (\eta +1)^2 \xi\eta\\
    &\quad +4 \pi  (\eta +1)^2  (4 \xi
   +1) \phi'^2 r^2+16 \pi  (\eta +1)^2  \xi  \phi' r \phi
   \end{split}
\]

\[
\begin{split}
C_1&=\phi' [-(\eta +1)^2+4 \pi  \xi r^2 \eta \left(2 m^2 \phi ^2+4 \rho \right)-2\pi  (\eta +1)  (-r^2 (m^2 (4 \xi +2) \phi ^2\\
&\quad+2 \rho )+24 \xi ^2 \phi^2+2 p r^2)]+(\eta +1) r \phi  \left(m^2 \left(-\eta +1\right)+8 \pi  \xi  \left(3 p-\rho \right)\right)
\end{split}
\]

\[
    C_0=8 \pi   \xi  \phi'^3 r^2+8 \pi   \xi  \phi'^2 r \phi \left(6 \eta  \xi +\eta -6 \xi +1\right)-(\eta +1) \phi' \left(\eta +6\xi\eta+1\right)
\]

\end{appendices}

\bibliographystyle{apsrev4-2}
\bibliography{references}

@article{Oertel2015, title={Hyperons in neutron star matter within relativistic mean-field models}, volume={42}, ISSN={0954-3899}, DOI={10.1088/0954-3899/42/7/075202}, number={7}, journal={Journal of Physics G: Nuclear and Particle Physics}, publisher={IOP Publishing}, author={Oertel, M. and Providência, C. and Gulminelli, F. and Raduta, Ad R.}, year={2015}, month={june}, pages={075202}, language={en} }

@article{MorisakiSuyama2017, title={Spontaneous scalarization with an extremely massive field and heavy neutron stars}, volume={96}, ISSN={2470-0010, 2470-0029}, DOI={10.1103/PhysRevD.96.084026}, number={8}, journal={Physical Review D}, author={Morisaki, Soichiro and Suyama, Teruaki}, year={2017}, month=oct, pages={084026} }

@article{Chen_Suyama_Yokoyama_2015, title={Spontaneous scalarization: asymmetron as dark matter}, url={https://arxiv.org/abs/1508.01384v1}, DOI={10.1103/PhysRevD.92.124016}, journal={arXiv.org}, author={Chen, Pisin and Suyama, Teruaki and Yokoyama, Jun’ichi}, year={2015}, month=aug, language={en} }

@article{DEF1993,
  title = {Nonperturbative strong-field effects in tensor-scalar theories of gravitation},
  author = {Damour, Thibault and Esposito-Far\`ese, Gilles},
  journal = {Phys. Rev. Lett.},
  volume = {70},
  issue = {15},
  pages = {2220--2223},
  numpages = {0},
  year = {1993},
  month = {Apr},
  publisher = {American Physical Society},
  doi = {10.1103/PhysRevLett.70.2220},
  url = {https://link.aps.org/doi/10.1103/PhysRevLett.70.2220}
}

@article{DEF1996,
  title = {Tensor-scalar gravity and binary-pulsar experiments},
  author = {Damour, Thibault and Esposito-Far\`ese, Gilles},
  journal = {Phys. Rev. D},
  volume = {54},
  issue = {2},
  pages = {1474--1491},
  numpages = {0},
  year = {1996},
  month = {Jul},
  publisher = {American Physical Society},
  doi = {10.1103/PhysRevD.54.1474},
  url = {https://link.aps.org/doi/10.1103/PhysRevD.54.1474}
}

@article{Demorest_2010,
   title={A two-solar-mass neutron star measured using Shapiro delay},
   volume={467},
   ISSN={1476-4687},
   url={http://dx.doi.org/10.1038/nature09466},
   DOI={10.1038/nature09466},
   number={7319},
   journal={Nature},
   publisher={Springer Science and Business Media LLC},
   author={Demorest, P. B. and Pennucci, T. and Ransom, S. M. and Roberts, M. S. E. and Hessels, J. W. T.},
   year={2010},
   month=oct, pages={1081–1083} }

@article{Ramazano2016,
   title={Spontaneous scalarization with massive fields},
   volume={93},
   ISSN={2470-0029},
   url={http://dx.doi.org/10.1103/PhysRevD.93.064005},
   DOI={10.1103/physrevd.93.064005},
   number={6},
   journal={Physical Review D},
   publisher={American Physical Society (APS)},
   author={Ramazanoğlu, Fethi M. and Pretorius, Frans},
   year={2016},
   month=mar }

@article{Arapogulu2019,
  title = {Neutron star structure in the presence of nonminimally coupled scalar fields},
  author = {Arapo\ifmmode \breve{g}\else \u{g}\fi{}lu, A. Sava\ifmmode \mbox{\c{s}}\else \c{s}\fi{} and Ek\ifmmode \mbox{\c{s}}\else \c{s}\fi{}i, K. Yavuz and Y\"ukselci, A. Emrah},
  journal = {Phys. Rev. D},
  volume = {99},
  issue = {6},
  pages = {064055},
  numpages = {10},
  year = {2019},
  month = {Mar},
  publisher = {American Physical Society},
  doi = {10.1103/PhysRevD.99.064055},
  url = {https://link.aps.org/doi/10.1103/PhysRevD.99.064055}
}

@article{Antoniadis2013,
   title={A Massive Pulsar in a Compact Relativistic Binary},
   volume={340},
   ISSN={1095-9203},
   url={http://dx.doi.org/10.1126/science.1233232},
   DOI={10.1126/science.1233232},
   number={6131},
   journal={Science},
   publisher={American Association for the Advancement of Science (AAAS)},
   author={Antoniadis, John and Freire, Paulo C. C. and Wex, et al.},
   year={2013},
   month=apr }

@article{Hui:2021tkt,
    author = "Hui, Lam",
    title = "{Wave Dark Matter}",
    eprint = "2101.11735",
    archivePrefix = "arXiv",
    primaryClass = "astro-ph.CO",
    doi = "10.1146/annurev-astro-120920-010024",
    journal = "Ann. Rev. Astron. Astrophys.",
    volume = "59",
    pages = "247--289",
    year = "2021"
}

@article{Staykov2018, title={Static and slowly rotating neutron stars in scalar–tensor theory with self-interacting massive scalar field}, volume={78}, ISSN={1434-6052}, DOI={10.1140/epjc/s10052-018-6064-x}, number={7}, journal={The European Physical Journal C}, author={Staykov, Kalin V. and Popchev, Dimitar and Doneva, Daniela D. and Yazadjiev, Stoytcho S.}, year={2018}, month={july}, pages={586} }

@article{Odintsov2021, title={Neutron Stars in Scalar-tensor Gravity with Quartic Order Scalar Potential}, url={https://arxiv.org/abs/2104.01982v2}, DOI={10.1016/j.aop.2022.168839}, journal={arXiv.org}, author={Odintsov, S. D. and Oikonomou, V. K.}, year={2021}, month=apr, language={en} }

@article{Prakash1996, title={Composition and Structure of Protoneutron Stars}, url={https://arxiv.org/abs/nucl-th/9603042v1}, DOI={10.1016/S0370-1573(96)00023-3}, journal={arXiv.org}, author={Prakash, Madappa and Bombaci, Ignazio and Prakash, Manju and Ellis, Paul J. and Lattimer, James M. and Knorren, Roland}, year={1996}, month=mar, language={en} }

@article{Nishizaki2002, title={Hyperon-Mixed Neutron Star Matter and Neutron Stars}, volume={108}, ISSN={0033-068X, 1347-4081}, DOI={10.1143/PTP.108.703}, number={4}, journal={Progress of Theoretical Physics}, author={Nishizaki, S. and Yamamoto, Y. and Takatsuka, T.}, year={2002}, month=oct, pages={703–718}, language={en} }

@article{Oppenheimer1939, title={On Massive Neutron Cores}, volume={55}, rights={http://link.aps.org/licenses/aps-default-license}, ISSN={0031-899X}, DOI={10.1103/PhysRev.55.374}, number={4}, journal={Physical Review}, author={Oppenheimer, J. R. and Volkoff, G. M.}, year={1939}, month=feb, pages={374–381}, language={en} }

@article{Tolman1939, title={Static Solutions of Einstein’s Field Equations for Spheres of Fluid}, volume={55}, rights={http://link.aps.org/licenses/aps-default-license}, ISSN={0031-899X}, DOI={10.1103/PhysRev.55.364}, number={4}, journal={Physical Review}, author={Tolman, Richard C.}, year={1939}, month=feb, pages={364–373}, language={en} }

@article{Banik2014, title={NEW HYPERON EQUATIONS OF STATE FOR SUPERNOVAE AND NEUTRON STARS IN DENSITY-DEPENDENT HADRON FIELD THEORY}, volume={214}, ISSN={0067-0049}, DOI={10.1088/0067-0049/214/2/22}, journal={The Astrophysical Journal Supplement Series}, publisher={IOP Publishing}, author={Banik, Sarmistha and Hempel, Matthias and Bandyopadhyay, Debades}, year={2014}, month={sept}, pages={22}, language={en} }

@article{Konstantinou2022,
   title={Universal Relations for the Increase in the Mass and Radius of a Rotating Neutron Star},
   volume={934},
   ISSN={1538-4357},
   url={http://dx.doi.org/10.3847/1538-4357/ac7b86},
   DOI={10.3847/1538-4357/ac7b86},
   number={2},
   journal={The Astrophysical Journal},
   publisher={American Astronomical Society},
   author={Konstantinou, Andreas and Morsink, Sharon M.},
   year={2022},
   month=aug, pages={139} }

@article{Weissenborn2012,
  title = {Hyperons and massive neutron stars: Vector repulsion and SU(3) symmetry},
  author = {Weissenborn, S. and Chatterjee, D. and Schaffner-Bielich, J.},
  journal = {Phys. Rev. C},
  volume = {85},
  issue = {6},
  pages = {065802},
  numpages = {9},
  year = {2012},
  month = {Jun},
  publisher = {American Physical Society},
  doi = {10.1103/PhysRevC.85.065802},
  url = {https://link.aps.org/doi/10.1103/PhysRevC.85.065802}
}

@article{Vidana2022, title={Neutron stars and the hyperon puzzle}, volume={271}, url={https://inspirehep.net/literature/2605804}, DOI={10.1051/epjconf/202227109001}, journal={EPJ Web Conf.}, author={Vidaña, Isaac}, year={2022}, language={en} }

@article{Doneva:2022ewd,
    author = "Doneva, Daniela D. and Ramazano{\u{g}}lu, Fethi M. and Silva, Hector O. and Sotiriou, Thomas P. and Yazadjiev, Stoytcho S.",
    title = "{Spontaneous scalarization}",
    eprint = "2211.01766",
    archivePrefix = "arXiv",
    primaryClass = "gr-qc",
    doi = "10.1103/RevModPhys.96.015004",
    journal = "Rev. Mod. Phys.",
    volume = "96",
    number = "1",
    pages = "015004",
    year = "2024"
}

@article{Punturo2010, title={The Einstein Telescope: a third-generation gravitational wave observatory}, volume={27}, DOI={10.1088/0264-9381/27/19/194002}, number={19}, journal={Classical and Quantum Gravity}, author={Punturo, M and Abernathy, M and Acernese, et al.}, year={2010}, month={sept}, pages={194002} }

@article{Hinderer2008, title={Tidal Love Numbers of Neutron Stars}, volume={677}, ISSN={0004-637X}, DOI={10.1086/533487}, number={2}, journal={The Astrophysical Journal}, publisher={IOP Publishing}, author={Hinderer, Tanja}, year={2008}, month=apr, pages={1216}, language={en} }

@article{Creci2023, title={Tidal properties of neutron stars in scalar-tensor theories of gravity}, volume={108}, DOI={10.1103/PhysRevD.108.124073}, number={12}, journal={Physical Review D}, publisher={American Physical Society}, author={Creci, Gastón and Hinderer, Tanja and Steinhoff, Jan}, year={2023}, month=dec, pages={124073}, language={en} }

@article{Ignazio2016,
  author  = {Ignazio Bombaci},
  title   = {The Hyperon Puzzle in Neutron Stars},
  journal = {JPS Conference Proceedings},
  volume  = {17},
  pages   = {101002},
  year    = {2016},
  doi     = {10.7566/JPSCP.17.101002}
}

@article{Degollado2024, title={Dynamical transition to spontaneous scalarization in neutron stars: The massive scalar field scenario}, volume={110}, ISSN={2470-0010, 2470-0029}, DOI={10.1103/PhysRevD.110.084011}, note={arXiv:2407.08124 [gr-qc]}, number={8}, journal={Physical Review D}, author={Degollado, Juan Carlos and Ortiz, Néstor and Salgado, Marcelo}, year={2024}, month=oct, pages={084011} }

\end{document}